\documentclass[journal]{IEEEtran}
\usepackage{cite}

\usepackage{subfigure}
\usepackage{graphicx}
\usepackage{graphics}
\usepackage{amssymb}
\usepackage{color}
\usepackage{booktabs}
\usepackage{threeparttable}

\usepackage[colorlinks=true, citecolor=blue]{hyperref}

\ifCLASSINFOpdf
\else
\fi
\usepackage[cmex10]{amsmath}
\usepackage{array}
\usepackage{url}

\usepackage[switch,left]{lineno}
\usepackage{xcolor}

\begin{document}

\twocolumn

\title{A Review of Stop-and-Go Traffic Wave Suppression Strategies: Variable Speed Limit vs. Jam-Absorption Driving}

%
%
%

\author{Zhengbing~He, 
Jorge~Laval,
Yu~Han,
Andreas~Hegyi,
Ryosuke~Nishi,
Cathy~Wu
\thanks{
Z. He is with Laboratory for Information and Decision Systems (LIDS), Massachusetts Institute of Technology, Cambridge MA, USA ({\it he.zb@hotmail.com, hezb@mit.edu})

J. Laval is with School of Civil and Environmental Engineering at Georgia Institute of Technology, Atlanta GA, USA ({\it jorge.laval@ce.gatech.edu})

Y. Han is with School of Transportation, Southeast University, Nanjing, China ({\it yuhan@seu.edu.cn})

A. Hegyi is with the Department of Transport and Planning, Delft University of Technology, Delft, The Netherlands ({\it a.hegyi@tudelft.nl})

R. Nishi is with Department of Mechanical and Physical Engineering, Faculty of Engineering, Tottori University, Tottori, Japan ({\it nishi@tottori-u.ac.jp})

C. Wu is with Laboratory for Information and Decision Systems, Institute for Data, Systems, and Society, and Department of Civil and Environmental Engineering, Massachusetts Institute of Technology, Cambridge MA, USA ({\it cathywu@mit.edu})
}

\thanks{Corresponding to Z. He. Manuscript received April, 2025}}


%
%

{}
%



\maketitle
\begin{abstract}
The main form of freeway traffic congestion is the familiar stop-and-go wave, characterized by wide moving jams that propagate indefinitely upstream provided enough traffic demand. They cause severe, long-lasting adverse effects, such as reduced traffic efficiency, increased driving risks, and higher vehicle emissions.
This underscores the crucial importance of artificial intervention in the propagation of stop-and-go waves.
Over the past two decades, two prominent strategies for stop-and-go wave suppression have emerged: variable speed limit (VSL) and jam-absorption driving (JAD).
Although they share similar research motivations, objectives, and theoretical foundations, the development of these strategies has remained relatively disconnected.
To synthesize fragmented advances and drive the field forward, this paper first provides a comprehensive review of the achievements in the stop-and-go wave suppression-oriented VSL and JAD, respectively. It then focuses on bridging the two areas and identifying research opportunities from the following perspectives: fundamental diagrams, secondary waves, generalizability, traffic state estimation and prediction, robustness to randomness, scenarios for strategy validation, and field tests and practical deployment.
We expect that through this review, one area can effectively address its limitations by identifying and leveraging the strengths of the other, thus promoting the overall research goal of freeway stop-and-go wave suppression.

\end{abstract}


\section{Introduction}\label{sec:Intro}

\subsection{Stop-and-Go Waves on Freeways}

While driving through freeway stop-and-go traffic, vehicles alternate between accelerating and decelerating, resulting in adverse effects, such as reduced traffic efficiency and longer travel times \cite{Yildirimoglu2013,Zhang2017c},
significant fuel consumption and increased emissions \cite{Li2020,Jiang2025},
and a substantial increase in the risk of traffic accidents \cite{Wang2009,Wang2024}.

\begin{figure*}[ht]
    \centering
    \includegraphics[width=\linewidth]{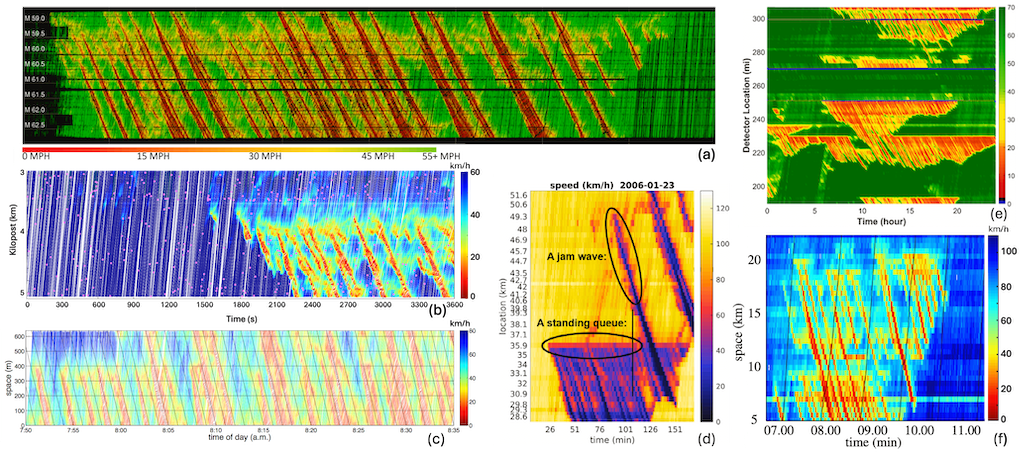}
    \caption{Global observations on freeway stop-and-go waves. 
    The striped low-speed regions propagating upstream against the direction of traffic flow are referred to as stop-and-go waves. 
    (a) Nashville, TN, USA: 6.7 km (sourced from \cite{Gloudemans2023});
    (b) Osaka City, Japan: 2 km (sourced from \cite{Dahiyal2020});
    (c) California, USA: 0.6 km (sourced from \cite{He2019});
    (d) Rotterdam, Netherlands: 23 km (sourced from \cite{VSL2022});
    (e) Florida, USA: 190 km (sourced from \cite{Staes2021});
    (f) London, UK: 18 km (sourced from \cite{Orosz2010}).}
    \label{fig:wave}
\end{figure*}

As shown in Figure \ref{fig:wave}, stop-and-go waves (also referred to as wide moving jams or traffic oscillations) are a common form of traffic congestion on freeways worldwide.
In the time-space diagrams of traffic speed, a stop-and-go wave appears as a striped low-speed region propagating upstream, against the direction of traffic flow.
Physically, on the freeway, a stop-and-go wave is essentially a backward-moving queue with an approximately constant queue length. 
As vehicles traverse these queues, they typically undergo a sequence of deceleration, low-speed cruising or even standstill, followed by acceleration.

The main features of stop-and-go waves are as follows. 
Stop-and-go waves propagate against the traffic direction at a speed of 10$\sim$20 km/h, i.e., the slope of the waves in the time-space diagram or backward-moving speed of the queues \cite{Mauch2002a,Laval2010,Zheng2011a,Jiang2014,He2015a,He2024}.
Depending on the speed of vehicles within a moving queue, the discharge rate (i.e., the outflow rate of the moving queue) can be up to 25\% lower than the capacity \cite{Yuan2014}.
It is common to observe that stop-and-go waves propagate upstream while maintaining their structure even tens of kilometers \cite{Orosz2010, Messner2012,Staes2021,Gloudemans2023}, severely impacting both traffic and the environment. 
As shown in Figure \ref{fig:wave}(e), some stop-and-go waves propagate upstream for as far as 50 km.



\subsection{Stop-and-Go Wave Suppression Strategies}

There are mainly two types of freeway traffic control strategies that were proposed to dissipate stop-and-go waves.
One strategy is to control traffic by disseminating variable speed limits (VSL), and the other is based on a dedicated vehicle to perform jam-absorption driving (JAD). 
VSL is typically disseminated through variable message signs (VMS) installed on gantries along the freeway. To ensure the effectiveness of an advanced VSL algorithm, a series of gantries is usually required, with the distance between two gantries commonly being about 500$\sim$1000 meters, as seen in the Netherlands \cite{VSL2010} and the United States \cite{Zhang2024}.
JAD is a driving strategy that guides a dedicated vehicle to perform ``slow-in" and ``fast-out" actions before being captured by a stop-and-go wave. Ideally, it can hold and guide all vehicles behind to approach an upcoming wave as planned, thereby achieving the goal of stop-and-go wave suppression.

The two types of strategies have many similarities, such as the goal, affected object, as well as theoretical and methodological questions (see Table \ref{tab:Similarities}). 
In particular, both share the same underlying logic, i.e., attempting to suppress stop-and-go waves by reducing the inflow.
In contrast, the primary differences arise from the control object: one strategy presents VSL via VMS, while the other employs a dedicated vehicle (see Table \ref{tab:Differences}).
As claimed by JAD literature, compared to VSL, JAD offers several advantages, such as requiring significantly less investment due to the absence of infrastructure construction and providing greater flexibility since VSL must be disseminated through fixed gantries and VMS.

\begin{table}[htbp]
\caption{Similarities between VSL and JAD.}\label{tab:Similarities}
\centering\footnotesize\renewcommand{\arraystretch}{1.2}
\setlength\tabcolsep{9pt}
\begin{tabular}{ll}
\toprule
Aspect & Details \\
\midrule
       Goal                       & Suppressing stop-and-go waves  \\
       Affected object            & Traffic flow   \\
       Underlying logic            & Reducing inflow   \\
       Theoretical question       & (1) Can stop-and-go waves be suppressed?   \\
                                  & (2) Will secondary waves be triggered?  \\
       Methodological question    & (1) When to activate the strategy?  \\
                                  & (2) How to implement the strategy?  \\        
\bottomrule
\end{tabular}
\end{table}

\begin{table}[htbp]
\caption{Differences between VSL and JAD.}\label{tab:Differences}
\centering\footnotesize\renewcommand{\arraystretch}{1.2}
\begin{tabular}{lll}
\toprule
Aspect & \multicolumn{2}{l}{{Details}}  \\
\cline{2-3} 
     & VSL & JAD\\
\midrule
       Main object of control         & Speed limit          & Vehicle  \\
       Feature of control object      & Continuous in time,  & Continuous in time, \vspace{-0.5mm}\\
                                      & discrete in space    & space and value \vspace{-0.5mm} \\       
                                      & and value            &   \\       
       Control source                 & Outside traffic$^\dagger$      &  Inside traffic$^\ddagger$  \\
       Difficulty of execution        & Low                  & High \\
       Flexibility of execution       & Low                  & High  \\
       Flexibility of deployment      & Low                  & High \\
       Cost of deployment             & High                 & Low \\
\bottomrule
\end{tabular}
\begin{tablenotes}\footnotesize
\item[1]  \hspace{-3mm} $\dagger$ VSL acts on the traffic flow from outside the flow itself.\\
\item[2]  \hspace{-3mm} $\ddagger$ The JAD vehicle acts within the traffic flow as an integral part of it.
\end{tablenotes}

\end{table}

\subsection{Motivation}\label{sec:Motivation}

Despite many similarities, particularly in goals, underlying logic, and research questions, a disconnection between the two areas is observed, as evidenced by the following two examples.

\begin{figure}[htbp]
    \centering
    \includegraphics[width=\linewidth]{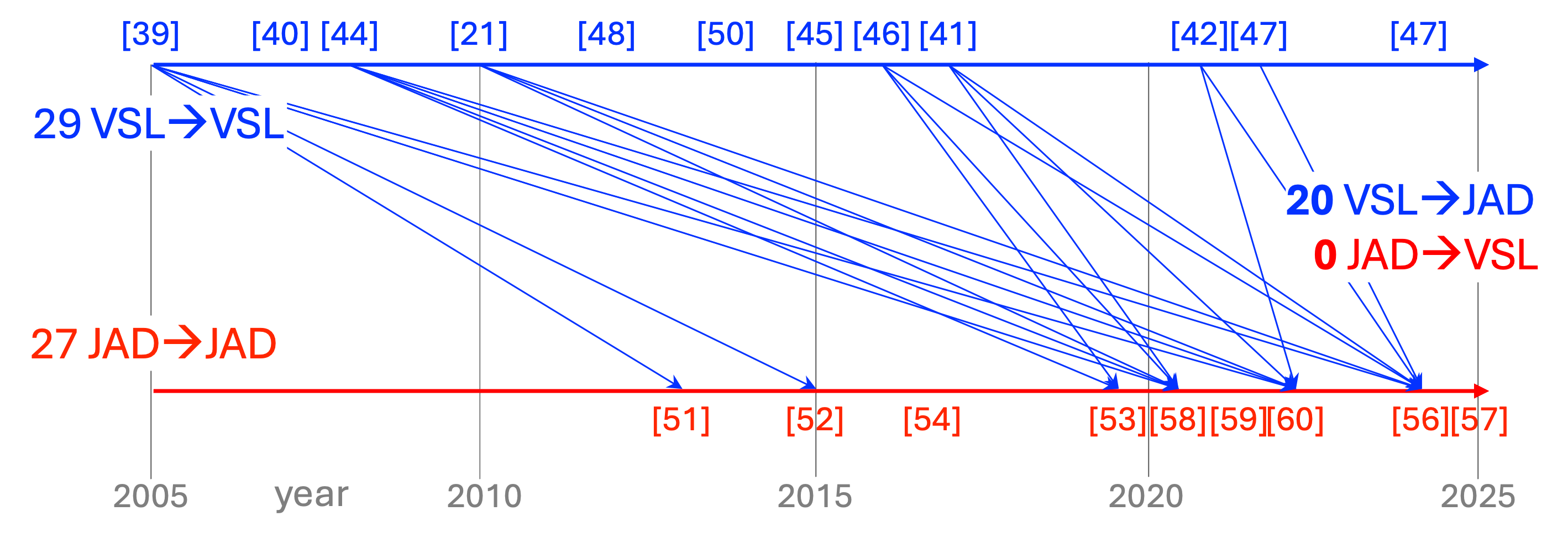}
    \caption{Citations among the selected VSL and JAD papers. 
    The VSL papers cite other VSL papers 29 times, and the JAD papers cite other JAD papers 27 times. While the JAD papers cite VSL papers 20 times, no VSL papers have cited the JAD papers.}
    \label{fig:citation}
\end{figure}

\begin{itemize}

    \item Among the selected VSL and JAD papers identified as representative works, none of the VSL papers cite JAD studies, even though, as we will show later, a trend toward incorporating connected and automated vehicles (CAVs) into VSL is observed.
    In contrast, JAD papers tend to reference VSL works, albeit not comprehensively (Figure \ref{fig:citation}).
    
    \item In VSL papers the term ``shock wave" is predominantly used, whereas JAD papers more commonly use ``stop-and-go wave" or ``oscillation." Although these terms describe similar types of traffic jams, the differences in terminology may lead to difficulties in discovering each other’s work through search engines.

\end{itemize}

The disconnection results in many researchers and engineers, including both authors and reviewers, lacking a comprehensive understanding of the overall progress in freeway traffic stop-and-go wave suppression, thereby hindering them from building their research upon existing findings.
In the meantime, to the best of our knowledge, no paper or review has specifically addressed or focused on these issues (one may refer to review papers such as \cite{Papageorgiou2003,Khondaker2015,Kusic2020,Vrbanic2021,Yu2021,Han2023}).
Therefore, a review paper that effectively structures the current research achievements and inspires future work is urgently needed.

\subsection{Contribution}

To fulfill the above motivations, this paper systematically reviews the current progress in strategies proposed to suppress freeway stop-and-go waves, exclusively through two techniques: VSL and JAD. 
We demonstrate that, despite the many similarities (Table \ref{tab:Similarities}), the two areas have developed distinct research focuses. 
For instance, VSL studies are strong in theory-based modeling but often lack a thorough consideration of secondary waves that may be triggered by the reduction of speed. 
In contrast, JAD studies are focused more on the microscopic perspective but lack a solid theoretical foundation and field test.
By leveraging the strengths of one area, the other can effectively address its limitations, thus promoting the overall research goal of stop-and-go wave suppression.

Moreover, we identify key research gaps in both areas and highlight future opportunities, drawing not only on foundational traffic flow theories and advanced artificial intelligence (AI) techniques but also by bridging the development paths of VSL and JAD in a unified framework.
In searching and reviewing related papers, we did find that some studies were not developing along sound directions. We, therefore, hope this review not only inspires future work but also helps avoid inappropriate assumptions and methods in the future.
Meanwhile, this review would also be beneficial for researchers in neighboring fields, such as CAV-based traffic control, ramp metering, and even eco-driving and traffic signal control at intersections.

\subsection{Scope}

We focus solely on the stop-and-go wave \textit{suppression} problem, in which stop-and-go waves have already \textit{fully developed}. 
Due to the motivations presented in Section \ref{sec:Motivation}, we concentrate exclusively on JAD and VSL strategies, as well as their comparison. The following topics are excluded to make this review paper a more specialized one.

\begin{itemize}

    \item  We do not discuss speed homogenization-based congestion mitigation \cite{Stern2018,Wang2022Optimal,Wu2022,Wang2023,Hayat2025,Lee2025}.  
    A distinguishing difference between JAD and speed homogenization might be the penetration rate of the controlled vehicles. JAD is expected to resolve the stop-and-go wave with only several controlled vehicles (usually, only one vehicle). 
    In contrast, speed homogenization commonly needs to control more vehicles. For instance, 4.5\% in the ring experiment in \cite{Stern2018,Wu2022} or a platoon of CAV in \cite{Wang2023}.
    
    \item We exclude studies that aim to alleviate traffic jams by addressing the bottlenecks that trigger them, such as freeway ramps \cite{Chen2014b,Hyun2020} and lane drops \cite{Du2019,Zhang2023}.




    \item We primarily focus on works published after 2003, as one may refer to the review paper \cite{Papageorgiou2003} for earlier freeway control strategies. Actually, prior to 2005, the stop-and-go wave suppression problem received little attention or research.
    
\end{itemize}

In addition, although some JAD studies position themselves in the framework of CAV technologies, we argue that we may not have to rely on the technologies when implementing JAD.
In Section \ref{sec:Field}, we will show that the swerving behavior of a police car is practically observed on California freeways to suppress traffic at high speeds. This observation greatly expands the potential for practical applications of JAD, especially at present, and encourages us to position JAD beyond CAV-based strategies.

\subsection{Organization}
The rest of the paper is organized as follows.
Section \ref{sec:Strategy} provides a detailed overview of the developments in VSL and JAD strategies, introducing, organizing, and comparing representative studies.
Section \ref{sec:Bridging} bridges the two areas and identifies future research opportunities from the following perspectives: fundamental diagrams (FDs), secondary waves, generalizability of strategies, traffic state estimation and prediction, stochasticity, scenarios for strategy validation, and field tests and practical deployment.
Section \ref{sec:Conclusion} concludes the review by summarizing the main findings and key insights.

\section{Variable Speed Limit vs. Jam-Absorption Driving}\label{sec:Strategy}

This section presents the development in freeway stop-and-go wave suppression strategies through VSL and JAD. 
Key features of VSL and JAD studies, selected based on our review, are summarized in Table \ref{tab:Summary}, the details of which will not be reiterated individually in the main text unless they are particularly important and warrant specific discussion.

\begin{table*}[htbp]
\caption{Summary of the selected VSL and JAD strategies.}\label{tab:Summary}
\centering\footnotesize\renewcommand{\arraystretch}{1.3}
\begin{tabular}{p{0.5cm}p{0.7cm}p{0.8cm}p{0.7cm}p{1.8cm}p{5cm}p{3.8cm}p{1cm}}
\toprule
Type & Feature & Paper & Year & \multicolumn{2}{l}{{Modeling}} &  \multicolumn{2}{l}{{Validation}} \\
\cmidrule(lr){5-6} \cmidrule(lr){7-8} 
     & & & & Control Object & Details &  Model & Waves \\    
\midrule 
     VSL & MPC & \cite{VSL2005}  & 2005 & Traffic & METANET-based prediction                                   & METANET     & Single \\   
         & MPC & \cite{VSL2007}  & 2007 & Traffic & METANET-based prediction                                   & Paramics    & {\it Multiple} \\   
         & MPC & \cite{VSL2017}  & 2017 & Traffic & Linear MPC; Modified CTM                                               & METANET     & Single \\ 
         & MPC & \cite{VSL2021}  & 2021 & Traffic, {\it Vehicle} & Speed-based MPC with safety constraints. Upper: VSL; Lower: CAV & VISSIM, IDM+ \cite{Wouter2012} & Single\vspace{1mm} \\  
         & KWT & \cite{VSL2008} & 2008 & Traffic & SPECIALIST (\reflectbox{$\lambda$}-shape FD)                                  & METANET      & Single \\   
         & KWT & \cite{VSL2010} & 2010 & Traffic & SPECIALIST                                                          & Real-world Test   & Single \\   
         & KWT & \cite{VSL2015} & 2015 & Traffic & SPECIALIST, Ramp metering                                          & VISSIM       & Single \\ 
         & KWT & \cite{VSL2016} & 2016 & Traffic, {\it Vehicle}   & SPECIALIST, Upper: VSL; Lower: CAV                 & IDM+         & Single \\ 
         & KWT & \cite{VSL2024} & 2024 & {\it Vehicle} & SPECIALIST (Triangular FD)                                            & SUMO (3 lanes)  & Single\vspace{1mm} \\  
         & TPTT & \cite{VSL2012} & 2012 & Traffic & New traffic patterns found        & KKS CA \cite{Kerner2011b} & Single \\   
         & TPTT & \cite{VSL2014} & 2014 & Traffic & Principle-based VSL coordination & KKS CA                    & Single\vspace{1mm} \\            
         & RL  & \cite{VSL2024} & 2022 & Traffic & Model free: offline training + online control &  METANET, Modified CTM \cite{VSL2017} & Single \\     
\midrule          
     JAD & Analysis     & \cite{JAD2013}  & 2013 & Vehicle & Concave, convex, and triangular FDs      & \multicolumn{2}{l}{N/A} \\   
         & Analysis     & \cite{JAD2015}  & 2015 & Vehicle & Helly CF model, IDM                                    & \multicolumn{2}{l}{N/A} \\   
         & Analysis     & \cite{JAD2020b} & 2020 & Vehicle & IDM                                      & \multicolumn{2}{l}{N/A} \vspace{1mm}  \\ 
         & Rule     & \cite{JAD2017}  & 2017 & Vehicle & Newell CF theory \cite{Newell2002}                        & $k$NN CF model + US data \cite{He2015a} & {\it Multiple} \\   
         & Rule     & \cite{JAD2024b} & 2024 & Vehicle & Targeting multiple waves; Real-time JAD speed modification    & IDM+                 & {\it Multiple} \vspace{1mm}  \\
         & MPC     & \cite{JAD2025} & 2025 & Vehicle & Stochastic CF model-based wave propagation estimation    & $k$NN CF model + Japanese data         & {\it Multiple} \vspace{1mm}  \\
         & Safety      & \cite{JAD2020}  & 2020 & Vehicle & Platoon; Wavelet Transform                                  & IDM    & Single  \\            
         & Safety      & \cite{JAD2021}  & 2021 & Vehicle & Platoon                                      & IDM    & Single  \\   
         & Safety      & \cite{JAD2021b} & 2021 & Vehicle & Platoon; Optimal control; Capacity drop     & IDM+   & Single\vspace{1mm} \\                     
            
\bottomrule
\end{tabular}
\end{table*}

\subsection{Variable Speed Limit}

Primarily, VSL strategies have been developed along two main threads: model predictive control (MPC)-based strategies and kinematic wave theory (KWT)-based strategies.
In the early stage, the MPC was built upon METANET \cite{VSL2005}, which is a second-order, deterministic, cell-based macroscopic traffic flow model \cite{Papageorgiou1990,Messner2012,Wang2022}. 
The strategies divide a freeway stretch into a series of cells (i.e., segments). 
Each cell, approximately 300 to 1000 meters in length, corresponds to a VSL.
The MPC-based strategy optimally coordinates VSL at different segments with the objective of minimizing total travel time. It is known that the effectiveness of MPC is directly determined by the accuracy of the predictive model used, turning out that the representativeness of METANET is the key to the effectiveness of the MPC-based strategies.
Initially, \cite{VSL2005} tested the METANET-based MPC solely within a METANET-based simulation, which is equivalent to assuming perfect prediction.
To relax the assumption, \cite{VSL2007} conducted a more rigorous and practical validation using the microscopic traffic flow simulation software Paramics 5.1, a commercial tool developed by Quadstone.
Further, to address the high computational burden associated with nonlinear and non-convex MPC, \cite{VSL2017} proposed a linear MPC-based strategy with a modified first-order cell transmission model (CTM), achieving similar optimization results with significantly improved efficiency.
More recently, \cite{VSL2021} introduced a linear Lagrangian MPC, which utilizes average vehicle speeds as decision variables rather than flow, as in previous MPC-based strategies. This modification bridges traffic-oriented control and vehicle-oriented control, enhancing the flexibility of the overall framework of the stop-and-go wave suppression.

\begin{figure}[htbp]
    \centering
    \includegraphics[width=\linewidth]{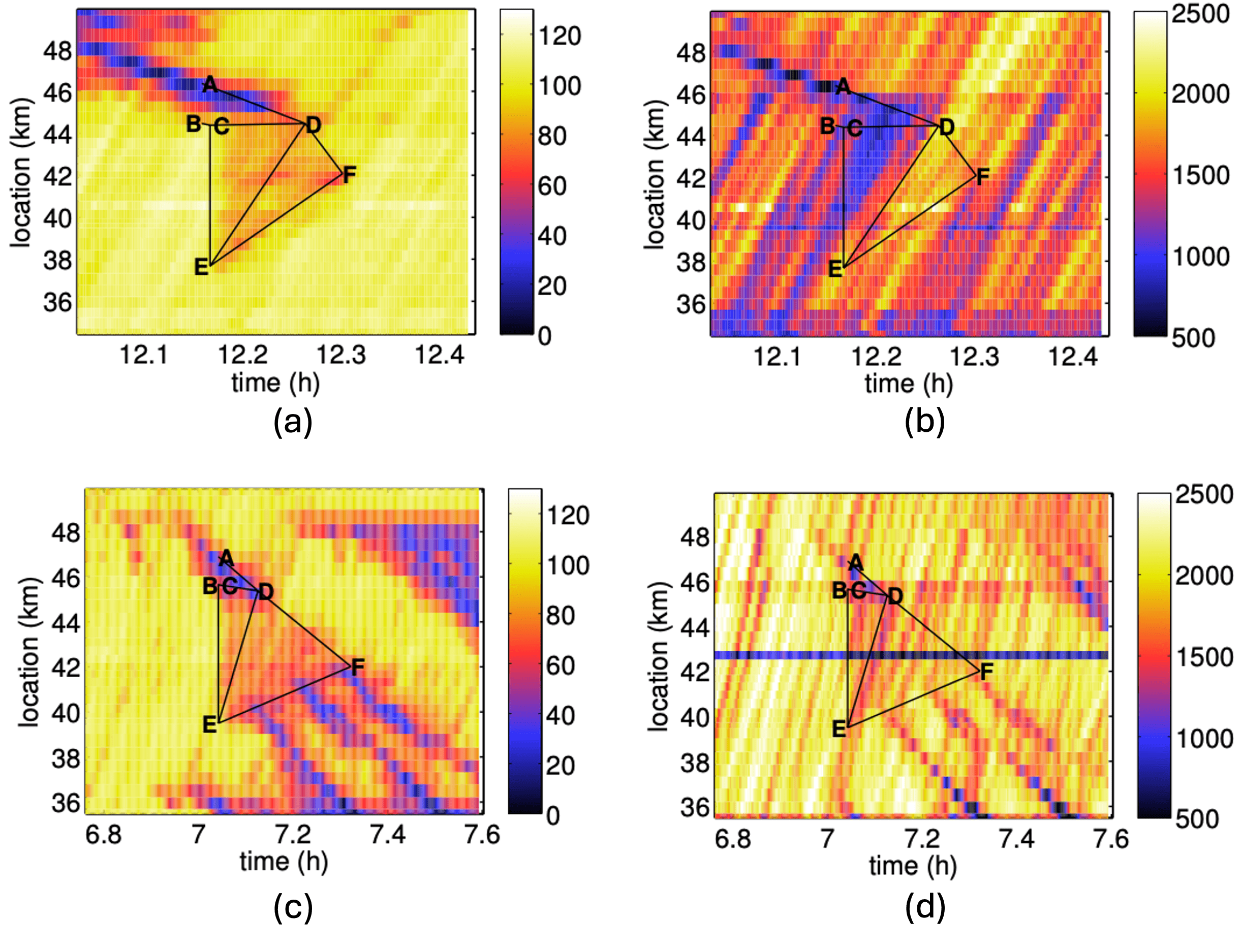}
    \caption{Examples of the field test results in \cite{VSL2010}. 
    (a)(b) A successfully resolved stop-and-go wave: speed (km/h) and flow (veh/h/lane);
    (c)(d) A case where secondary waves were triggered: speed (km/h) and flow (veh/h/lane).}
    \label{fig:field}
\end{figure}

Another thread is KWT-based VSL strategies.
\cite{VSL2008} proposed the seminal algorithm (named SPECIALIST) by conducting a theoretical analysis using KWT on an FD with a capacity drop (i.e., \reflectbox{$\lambda$}-shape FD). The opportunity to suppress a stop-and-go wave arises from the empirical observation of an outflow reduction of up to 30\% associated with a stop-and-go wave \cite{Kerner1996}.
Later, in 2010, \cite{VSL2010} conducted a field test of SPECIALIST to suppress stop-and-go waves over a 14 km stretch of the Dutch A12 freeway, which, to date, may be the only real-world test of such stop-and-go wave suppression strategies.
Although the effectiveness of the algorithm was well demonstrated (Figure \ref{fig:field}(a)(b)), it was found that the algorithm was frequently activated for other types of jams rather than stop-and-go waves. 
Notably, it was reported that secondary waves were triggered by the newly imposed lower speed limit (Figure \ref{fig:field}(c)(d)). 
\cite{VSL2015} introduced ramp metering to the SPECIALIST framework and proposed a coordinated strategy with mainstream control and ramp metering.
To take advantage of CAVs, \cite{VSL2016} introduced a bi-level framework to combine VSL and CAVs and enhance the suppression (actually earlier than \cite{VSL2021}, where a similar bi-level framework was adopted).
To deliver VSL directly to vehicles, the latest work reformulated a \reflectbox{$\lambda$}-shape FD into a triangular form and proposed a Lagrangian strategy capable of generating VSL suitable for implementation in connected vehicles \cite{VSL2024}.
\textit{It is important to note that} SPECIALIST and its variants are built upon FDs that exhibit capacity drop, meaning they are only applicable to stop-and-go waves where capacity drop occurs. As we will elaborate in Section \ref{sec:FD}, only \textit{isolated} stop-and-go waves—with high-density free-flow conditions both upstream and downstream—may trigger capacity drop. An example is the wave labeled “A jam wave” in Figure \ref{fig:wave}(d), which differs from most of the waves shown in Figure \ref{fig:wave}.
In contrast, although MPC-based strategies were also proposed to dissipate the same type of waves, their underlying mechanisms and optimization objectives (i.e., minimizing total time spent) suggest that they may be effective for a broader range of wave types, provided that the control resolution is increased.
As we will show in what follows, JAD does not specifically target this type of waves.




As a theory grounded in empirical observation, the three-phase traffic theory (TPTT) \cite{Kerner2011} was also employed to investigate VSL strategies \cite{VSL2012,VSL2014}. As expected, the new traffic flow model has led to new strategies and findings, such as the traffic evolution different from the seminal one in \cite{VSL2008}, although the TPTT-based strategy itself has not yet been empirically tested.


More recently, a model-free reinforcement learning (RL)-based VSL strategy was proposed \cite{VSL2022}, distinguishing itself from the previous model-based approaches. It employs an iterative training framework, where the optimal control policy is updated by exploring new control actions both online and offline in each iteration, thus avoiding the reliance on an RL model on a simulator.
Compared to other fields in transportation, the adoption of AI technology in stop-and-go wave suppression is relatively slow.

Among these VSL studies, a trend is observed, i.e., vehicle control, as a supplement to the control of traffic, has attracted more attention in recent years. Due to the rapid development of CAV and vehicle-to-infrastructure (V2I) technologies, the utilization of CAVs has been gradually increasing, implying that integrating traffic-oriented VSL and vehicle-oriented JAD could better facilitate the advancement of stop-and-go wave suppression.

Even though secondary waves were observed in the real world as early as 2010 \cite{VSL2010}, subsequent VSL studies have not paid much attention to this issue. This contrasts sharply with JAD strategies, which have extensively discussed the secondary wave issue, as we will show in Section \ref{sec:JAD}.
Another reason why the secondary wave issue has not received much attention may lie in the selection of simulation models. Some microscopic models and macroscopic models may be too stable to replicate traffic breakdowns caused by, for instance, the imperfect driving behavior of human drivers. 

\subsection{Jam-Absorption Driving}\label{sec:JAD}

The concept of JAD was initially introduced in a manner similar to a popular science article, as discussed by Beaty on a website \cite{Beaty1998, Beaty2013}. It was not until 2013 that \cite{JAD2013} sought to systematically examine and analyze its potential in a scientifically rigorous manner.
From the very beginning, the research questions were framed around two key questions: (1) Can JAD suppress stop-and-go waves? (2) Will secondary waves be triggered?
Centered around these two questions, \cite{JAD2013} conducted a theoretical analysis based on KWT, focusing more on the impact of FD shapes (i.e., concave, convex, and triangular FDs) rather than on a capacity drop, which was the main focus of KWT-based VSL studies \cite{VSL2008}.
It concluded that JAD is effective only when the inflow is lower than the capacity. Unfortunately, this condition does not support the self-sustainability of a stop-and-go wave that has been well demonstrated by the empirical observations in Figure \ref{fig:wave}. We will discuss this in detail in Section \ref{sec:FD}.
Later, \cite{JAD2015,JAD2020b} extended the analysis from a microscopic perspective by employing the Helly CF model and the intelligent driver model (IDM). The explicit formulations of vehicle dynamics in these CF models make the analyses traceable. However, it is important to be aware that the conclusions rely on the strong assumptions inherent in these models.

\begin{figure}[htbp]
    \centering
    \includegraphics[width=\linewidth]{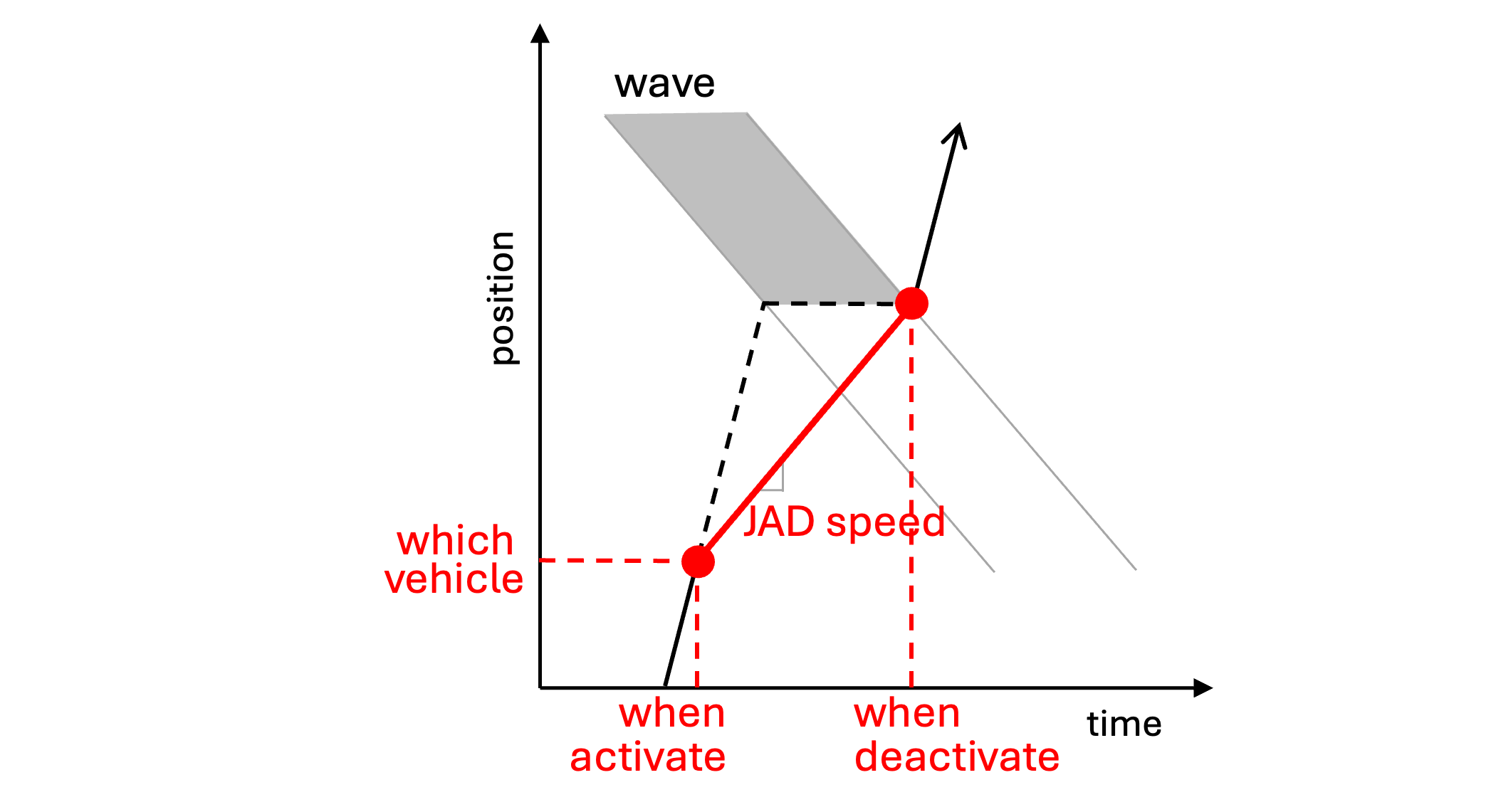}
    \caption{Research questions for a JAD strategy.}
    \label{fig:JAD}
\end{figure}

Since \cite{JAD2017} in 2017, more practical, implementation-oriented JAD strategies have been proposed. 
For a JAD strategy, given a predicted stop-and-go wave propagation track, the core tasks are to determine (1) when to activate and deactivate a JAD vehicle (2) which vehicle is appropriate to implement JAD (Figure \ref{fig:JAD}). The information is directly related to JAD speed.
\cite{JAD2017} predicted the propagation of a stop-and-go wave using the Newell CF theory \cite{Newell2002} and determined the JAD vehicle based on the principle of activating a JAD vehicle as early as possible. 
It treated the JAD speed as an exogenous parameter but demonstrated through simulation that a higher JAD speed leads to more effective wave suppression.
As a related study on determining when to activate JAD and which vehicle should be selected, \cite{Jerath2015} introduced the concept of an ``influential subspace”, which was defined as the region within which a vehicle can dissipate a single wave within a predetermined time period.
Recently, \cite{JAD2021b} introduced the concept of a capacity drop to the JAD study and proposed an optimal control-based JAD strategy that jointly optimizes traffic efficiency and safety by controlling the speed of JAD vehicles.
\cite{JAD2024b} targeted multiple wave suppression in one shot and designed a comprehensive set of steps to suppress the diverse evolutions of multiple stop-and-go waves.
Other remarkable efforts include that \cite{JAD2020} estimated the propagation of a stop-and-go wave using wavelet transform (WT), and \cite{JAD2020,JAD2021,JAD2021b} considered traffic efficiency and safety at the same time.
The latest work randomized the parameters of the Newell CF model to estimate the stop-and-go wave propagation and incorporated the JAD task into an MPC-based vehicle control framework \cite{JAD2025}.

\begin{figure}[htbp]
    \centering
    \includegraphics[width=\linewidth]{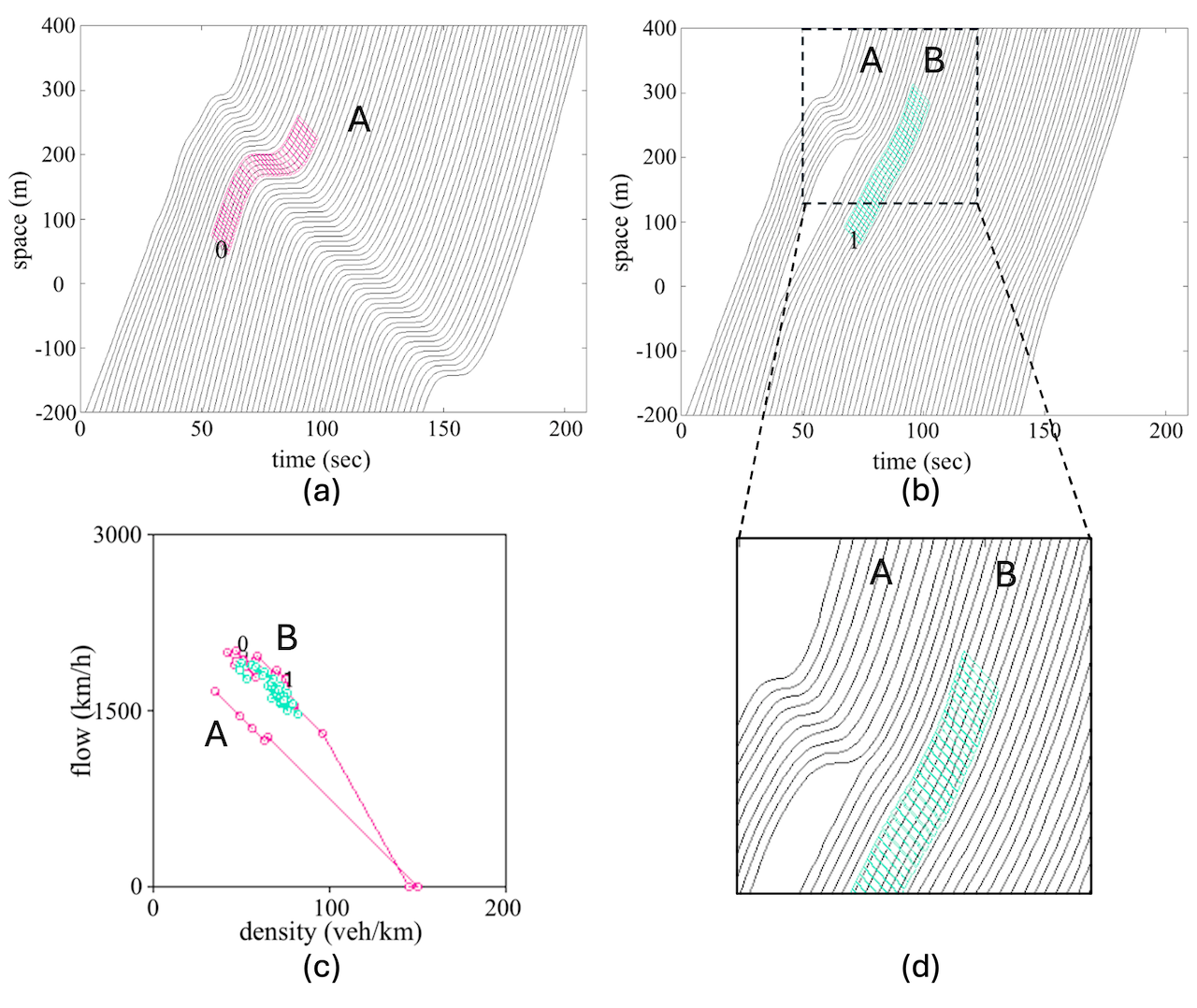}
    \caption{A successful JAD implementation in \cite{JAD2017}.
    (a) Before JAD; (b) After JAD; (c) Comparison in a flow-density plot; (d) An enlarged view of the dashed region in (b).}
    \label{fig:opportunity}
\end{figure}

A closer look at the JAD-involved simulation reveals how a wave is suppressed by a JAD vehicle: as shown in Figure \ref{fig:opportunity}, compared to the queue discharge rate at A, JAD results in a higher flow at B, even though the speed at B cannot exceed that at A due to the constraint imposed by the vehicles at A.

Up until now, according to the latest publication \cite{JAD2025}, the issue that secondary waves are triggered remains unresolved.
The newly triggered secondary waves may largely contribute to increased travel time when absorbing dense stop-and-go waves \cite{JAD2017}. 
Drawing on insights from VSL strategies, we recommend incorporating the empirically observed outflow reduction caused by stop-and-go waves and the potential benefits of mitigating this phenomenon into the modeling process, as it may serve as a key remedy for improving traffic efficiency.
As shown in \cite{Yuan2014}, the queue discharge rate reduction is positively and linearly related to the average speed of vehicles within a moving queue, highlighting the potential opportunity of suppressing a wave, especially when the speed within the queue is low.

Another reason related to secondary waves might be the distance of JAD execution. 
As concluded in \cite{JAD2017}, to effectively absorb a stop-and-go wave, JAD should be implemented over at least several hundred meters to ensure that the speed reduction from free-flow speed to JAD speed is gradual and less abrupt.
In contrast, controlled segments in current VSL strategies are typically several kilometers long, sometimes exceeding even ten kilometers.
The setting of VSL strategies implies that the current distance of JAD execution may still be conservative, and the execution distance could be further extended.

It is worth noting that we have observed a growing trend in which a platoon of vehicles has been employed and controlled to implement JAD for suppressing stop-and-go waves \cite{JAD2020,JAD2021,JAD2021b,JAD2024b}.
However, this approach seems to deviate from the original motivation behind JAD strategies, which is to suppress stop-and-go waves using a {\it single (or very few) vehicle(s)}. 
We should be clearly aware that if we have the capability to \textit{stably and precisely} control a vehicle platoon, the platoon-control-based JAD strategies should be compared with high CAV penetration-based speed homogenization strategies \cite{Stern2018,Wang2022Optimal,Wu2022,Wang2023,Hayat2025,Lee2025}, rather than with the single-vehicle-based JAD strategies.
In Section \ref{sec:Field}, we will illustrate that single-vehicle-based JAD is actually quite feasible in practice.

Another concern with JAD is whether a single JAD vehicle can ensure stable traffic flow behind it. 
Simply slowing down a JAD vehicle may result in excessively high densities in its wake, potentially triggering a breakdown. 
In VSL, stabilization is achieved through a combination of controlled speed (enforced by the speed limits) and sufficiently low density (e.g., region CDEF in Figure \ref{fig:field}).
Therefore, a model of traffic instability urgently needs to be incorporated.

\section{VSL-JAD Bridging and Research Opportunities}\label{sec:Bridging}

\subsection{Fundamental Diagrams}\label{sec:FD}

Many existing strategies and analyses are based on the FDs or their Lagrangian variants (e.g., speed-spacing relationship). It can be asserted that the shape of the assumed FDs directly influences both the formulation of a strategy and its effectiveness.

\begin{figure}[htbp]
    \centering
    \includegraphics[width=\linewidth]{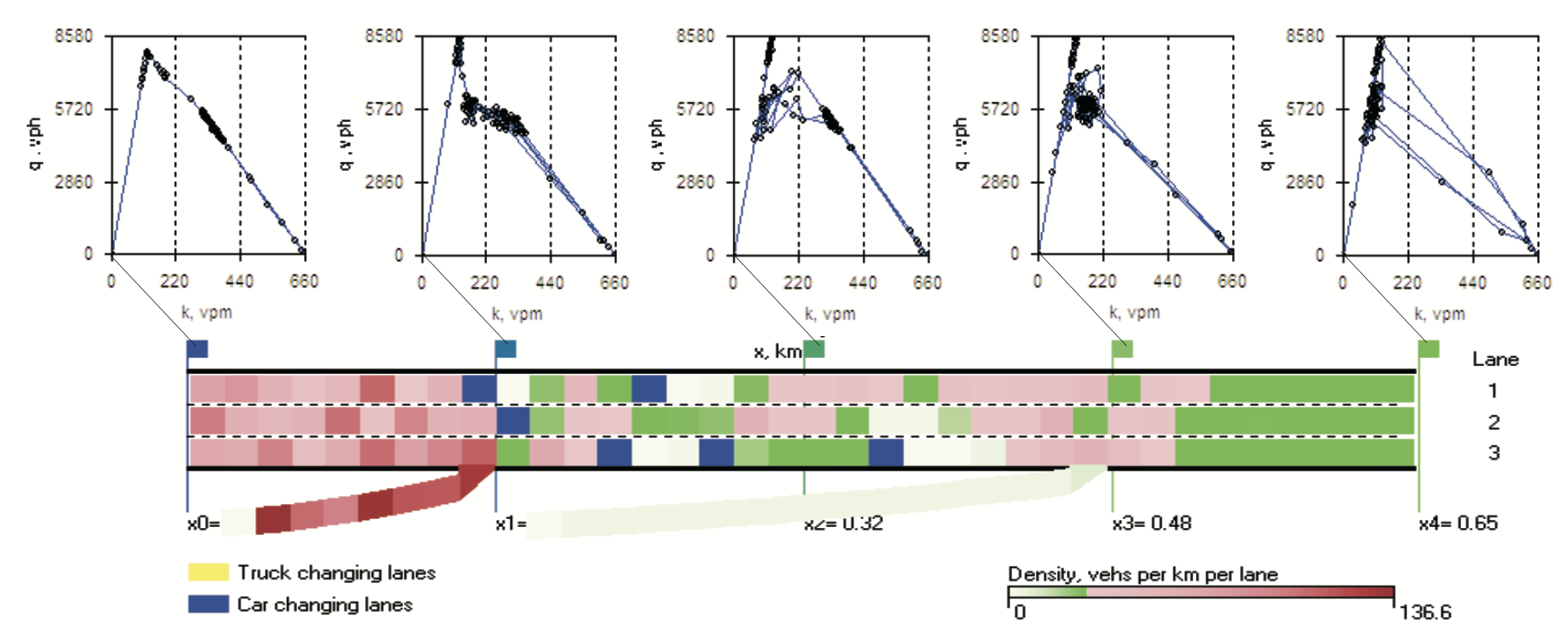}
    \caption{FDs associated with a bottleneck (sourced from \cite{Laval2007}). It shows that the upstream of a bottleneck corresponds to a triangular FD, while the downstream to a \reflectbox{$\lambda$}-shaped FD.}
    \label{fig:FD_shape}
\end{figure}

As shown in \cite{Laval2007}, the shape of an FD is closely related to its spatial relation to a bottleneck (Figure \ref{fig:FD_shape}). Specifically, an FD observed immediately downstream of a bottleneck typically exhibits a clear capacity drop (the second FD from the left in Figure \ref{fig:FD_shape}), whereas upstream of the bottleneck, the FD tends to approximate a triangular shape (the first FD from the left in Figure \ref{fig:FD_shape}).
Therefore, when assuming an FD, it is crucial to first determine the location or the type of the stop-and-go waves.

\begin{figure}[htbp]
    \centering
    \includegraphics[width=\linewidth]{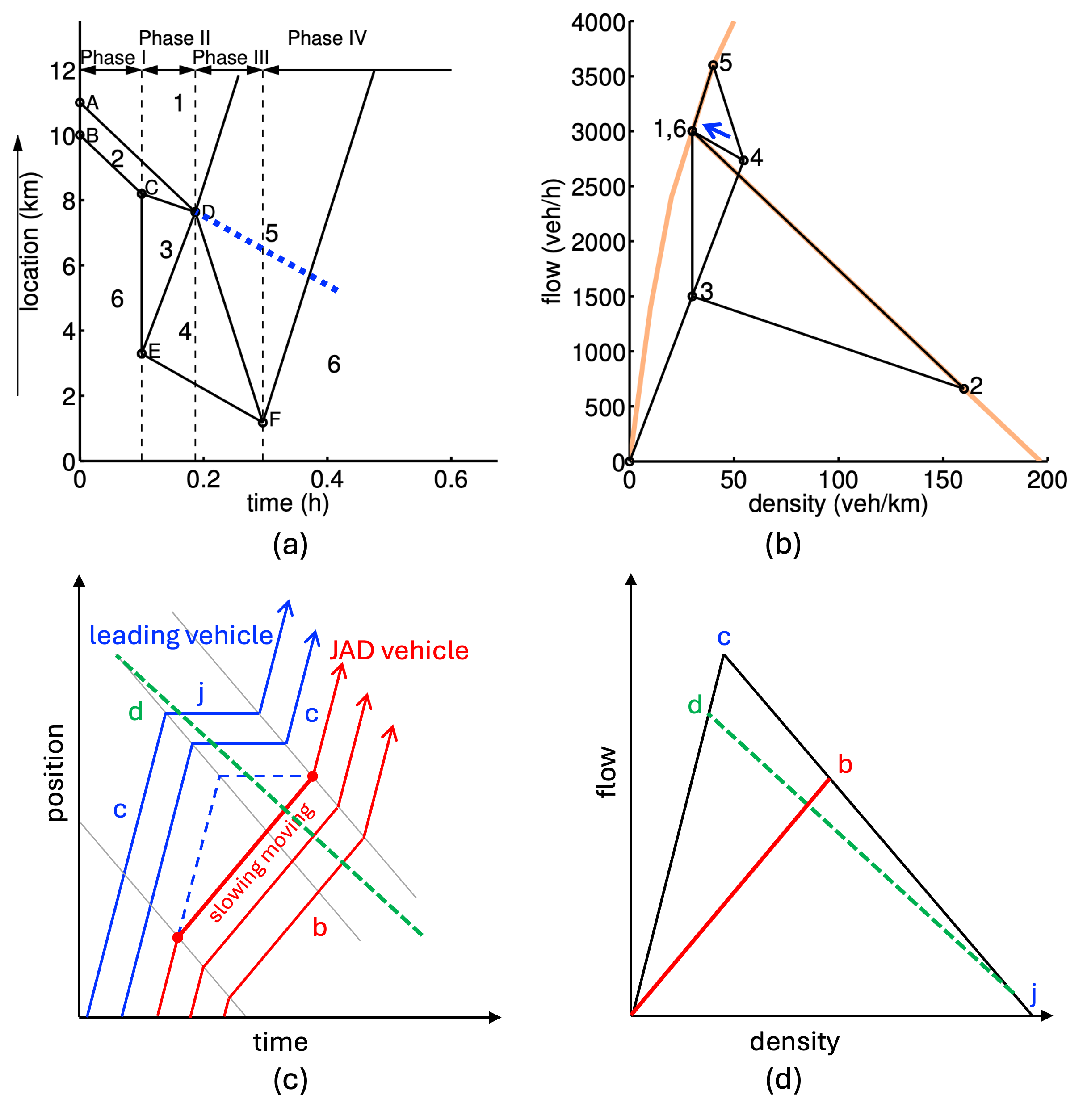}
    \caption{Schematic diagrams illustrating the traffic dynamics in the presence of suppression strategies and the assumed FDs. (a)(b) VSL (sourced from \cite{VSL2010}); (c)(d) JAD.}
    \label{fig:FD}
\end{figure}

For the SPECIALIST, it targets isolated stop-and-go waves (such as the one with the label ``A jam wave" in Figure \ref{fig:wave}(d)). 
Since the upstream and downstream traffic of such waves are in high-density free-flow conditions, the wave can be regarded as a (moving) bottleneck, and the associated FD may exhibit a capacity drop (Figure \ref{fig:FD}(b)).
This assumed \reflectbox{$\lambda$}-shape FD effectively resolves the contradiction between the widely observed constant length of a moving queue and the possibility of shrinking after imposing a lower speed limit.
More specifically, to be consistent with real-world observations (i.e., a constant length of a moving queue), the inflow and outflow of the queue must be equal (i.e., at states 1 and 6 in Figures \ref{fig:FD}(a) and (b)). 
Without capacity drop (i.e., no state 5), traffic will return from state 4 to state 1 (illustrated by the blue line and arrow in Figures \ref{fig:FD}(a) and (b), respectively), leading to the formation of another constant-length queue at state 4. 
More importantly, this queue at state 4 will not shrink and may continue to propagation upstream, although the density at state 4 is lower than the wave density at state 2.
It is noted that state 5 is highly unstable due to high flow \cite{Newell1962}.
Therefore, even though a stop-and-go wave is suppressed, as shown in Figure \ref{fig:FD}(a), it remains uncertain how long the high-flow region (state 5) can be sustained.

In contrast, JAD strategies target more broadly defined stop-and-go waves, rather than focusing solely on isolated waves. When multiple waves occur, as shown in the time-space diagrams in Figure \ref{fig:wave}, the traffic across the multiple and periodic stop-and-go waves is all in the congestion regime, and the corresponding FD is a triangle with no distinct capacity drop (similar to the first FD from the left in Figure \ref{fig:FD_shape}). Many JAD strategies do not make this distinction clearly, as the original JAD idea is simply to perform ``slow-in" and fast-out" actions.
For example, state b in Figure \ref{fig:FD}(c) is still self-sustained \cite{JAD2017}.
To achieve a self-shrinking JAD-impacted region with the triangular FD, \cite{JAD2013} concluded that an effective JAD requires the inflow to be lower than the capacity in the triangular FD (e.g., inflow at state d in Figure \ref{fig:FD}(d)).
It ensures the shrinking of the wave after JAD, but the condition implies that the wave, where JAD can take effect, can also shrink on its own (although it may take time), regardless of whether JAD control is applied (see the green line in Figure \ref{fig:FD}(c)).
More rigorous efforts are called for to establish a robust theoretical foundation grounded in empirical evidence.

Another two assumptions about FDs are also worth mentioning. 
The first is the use of three-phase FDs in \cite{VSL2012, VSL2014}, which leads to different conclusions compared to the ones based on the conventional two-phase FDs \cite{VSL2008}. 
Interested readers may look at Figure 6 in \cite{VSL2012}, which is quite similar to the empirical result shown in Figure \ref{fig:field}(a).
But we also remind that the TPTT may be explainable by the triangular FD-based KWT and lane-changing (LC) behavior \cite{Laval2007}.
The second one incorporates the state that is higher than the triangular-shaped states (i.e., state 5 in Figure \ref{fig:FD}(b)) into the triangle \cite{JAD2021b,VSL2024} or modifies the shape of the triangular FD to accommodate the higher capacity \cite{VSL2017,VSL2017b}, thereby achieving the goal of simplifying the modeling.

In addition, FD calibration is a more practical issue.
Both our practice and previous studies indicate that calibrating an FD in practice is challenging, mainly because of the lack of a unified method \cite{Bramich2022}.
An FD with a simpler expression can sometimes be even more difficult to calibrate.
For instance, for the triangular FD, identifying a well-accepted free-flow speed among the scattered empirical data points can be elusive, considering multi-lane traffic mixing with passenger cars and trucks \cite{Zheng2019}.
It particularly underscores the importance and urgent need for a robust approach that is less sensitive to such details.

\subsection{Secondary Waves}

There are three types of secondary waves.
\textit{Type-I secondary waves} are the ones that result directly from the implementation of speed reduction of upstream traffic.
In SPECIALIST, it corresponds to region EFD in Figure \ref{fig:field}(a) and Figure \ref{fig:FD}(a).
In JAD, it is region CEF in Figure 3 of \cite{JAD2013}.
Even though controlled by VSL, it is still unclear if the traffic in region EFD would breakdown under certain traffic conditions.

\textit{Type-II secondary waves} appear upstream of the controlled region (e.g., the upstream of line EF in Figure \ref{fig:field}(c)). These stop-and-go waves emerge upstream of the speed-limited vehicles, caused by vehicle accumulation at the boundary between a higher-flow upstream region and a lower-flow downstream region. In the SPECIALIST experiment, these new waves resulted from an overly high setting of the target density in region EFD in Figure \ref{fig:field}(a) and Figure \ref{fig:FD}(a). Since this specific traffic state—a particular combination of speed and density—had not previously been realized in practice, the exact stability threshold was unknown. After adjusting to a lower density value, these instabilities largely disappeared.

Unlike VSL, which controls a time-space region, JAD expects to suppress (or say, influence) a time-space region by directly controlling a time-space trajectory. As a result, many existing studies do not clearly distinguish between type-I and type-II secondary waves described above. The prevailing logic in JAD appears to be: dispatch a JAD vehicle, and if it fails to effectively absorb the secondary wave, dispatch another. Consequently, if the wave behind a JAD vehicle diminishes, it is considered that no secondary wave issue exists; if the wave persists or new waves emerge, it is taken as an indication of a secondary wave problem.

Note that most of the secondary wave-related issues mentioned in the previous sections belong to type-II, because a VSL is activated only when the type-I secondary wave is predicted to be finite. 
Inspired VSL, there is an urgent need to establish clear and reliable criteria for JAD to pre-determine whether activating a JAD vehicle can effectively absorb an upcoming stop-and-go wave. 
With such criteria, we can make more informed decisions about whether to deploy a JAD vehicle, rather than relying on ad-hoc decisions.

\textit{Type-III secondary waves} may occur downstream of the controlled region or trajectory, due to the higher outflow resulting from VSL or JAD interventions. However, since stop-and-go waves typically originate from infrastructural bottlenecks with limited capacity, if the intervention induces a forward wave with a flow close to or exceeding the bottleneck capacity, traffic may break down again downstream. 
A more thorough investigation of Type-III secondary waves is warranted, particularly with respect to their triggering mechanisms, contributing factors, and broader implications.


\subsection{Generalizability of Strategies}

\begin{figure}[htbp]
    \centering
    \includegraphics[width=\linewidth]{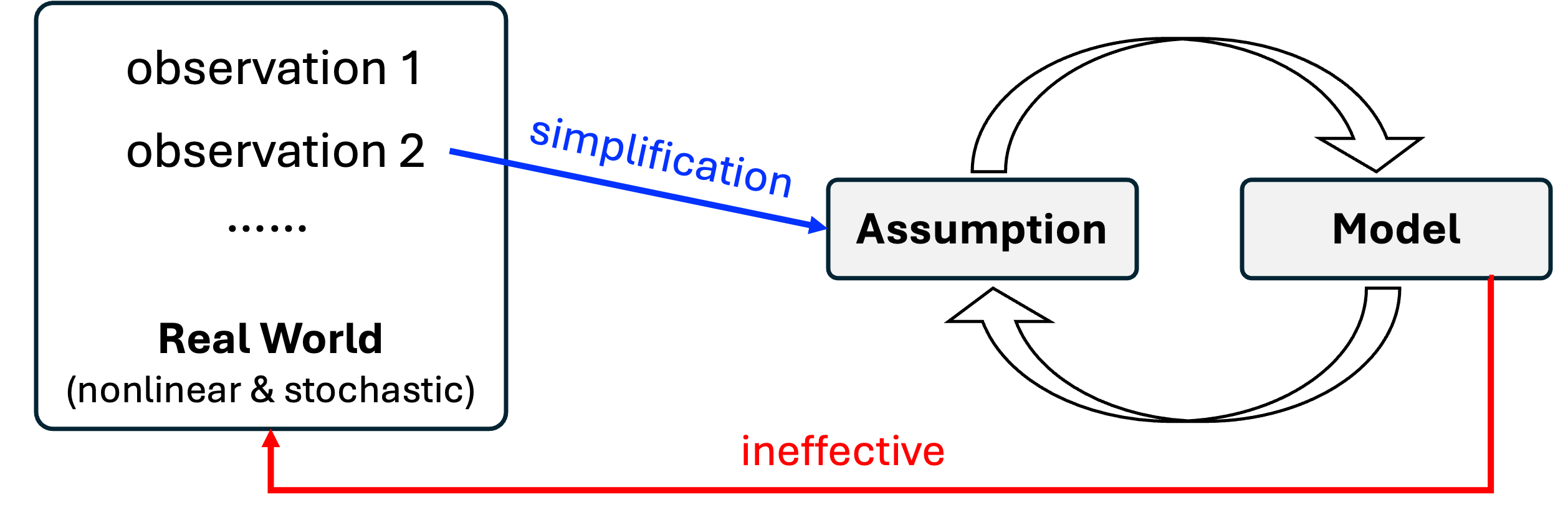}
    \caption{Complex real world vs. simplified assumptions and models.}
    \label{fig:consistency}
\end{figure}

Our researchers are focused on ensuring internal consistency in our assumptions and models (Figure \ref{fig:consistency}). In some cases, developing these models requires simplifying certain assumptions and scenarios, which, in turn, limits their effectiveness in real-world applications due to the inherently nonlinear and stochastic nature of real-world traffic systems \cite{Kerner2007}.
For instance, although the simulation-based findings (e.g., seven traffic states) in \cite{VSL2012} differ from the empirical ones (e.g., six traffic states) in \cite{VSL2010}, their work was also built upon a comprehensive TPTT, which is mainly rooted in empirical observations \cite{Kerner2011}.
It is difficult to dismiss them, as the world is diverse, and empirical observations from only a few locations cannot fully represent the complexity of reality.
Unlike other areas of transportation research, such as those centered on understanding traffic mechanisms or identifying general patterns, traffic control directly interacts with the system, placing exceptionally high demands on the robustness and adaptability of the models.

Recently, the rapidly-emerged Generative AI appears to be a potential tool to mitigate these challenges.
An insightful example might be PromptGAT \cite{Da2024}, which integrates Large Language Models (LLMs) to address the Sim-to-Real transfer performance gap in traffic signal control at intersections. 
The model utilizes the reasoning capabilities of LLMs to generate dynamic information based on specific conditions, such as weather, traffic status, and road types, which are then applied to enhance the action transformation process in RL models. 
This approach improves the applicability of simulation-based strategies in real-world environments, suggesting that LLMs can help theoretical models better adapt to real-world settings, which are often complex and challenging for humans.

Additionally, in some real-time applications, accurately determining or predicting which assumption best matches the actual situation is still challenging. 
This is not only due to the dynamic and stochastic nature of real-world conditions, but also because some information, particularly global information, cannot be obtained solely through observing local data.
As reported in \cite{VSL2010}, the proposed VSL strategy was activated in approximately half of the cases for isolated stop-and-go waves, and in the other half for different types of jams. It unveils the difficulty of identifying qualified stop-and-go waves in practice, i.e., distinguishing an isolated wave from multiple waves with certain criteria. 
Similarly, to quickly determine if a stop-and-go wave has consistently formed, \cite{JAD2017} had to propose a two-step estimation approach, which allows for an adjustment of the estimation after the initial judgment.

\subsection{Traffic State Estimation and Prediction}\label{sec:Prediction}

\begin{figure}[htbp]
    \centering
    \includegraphics[width=\linewidth]{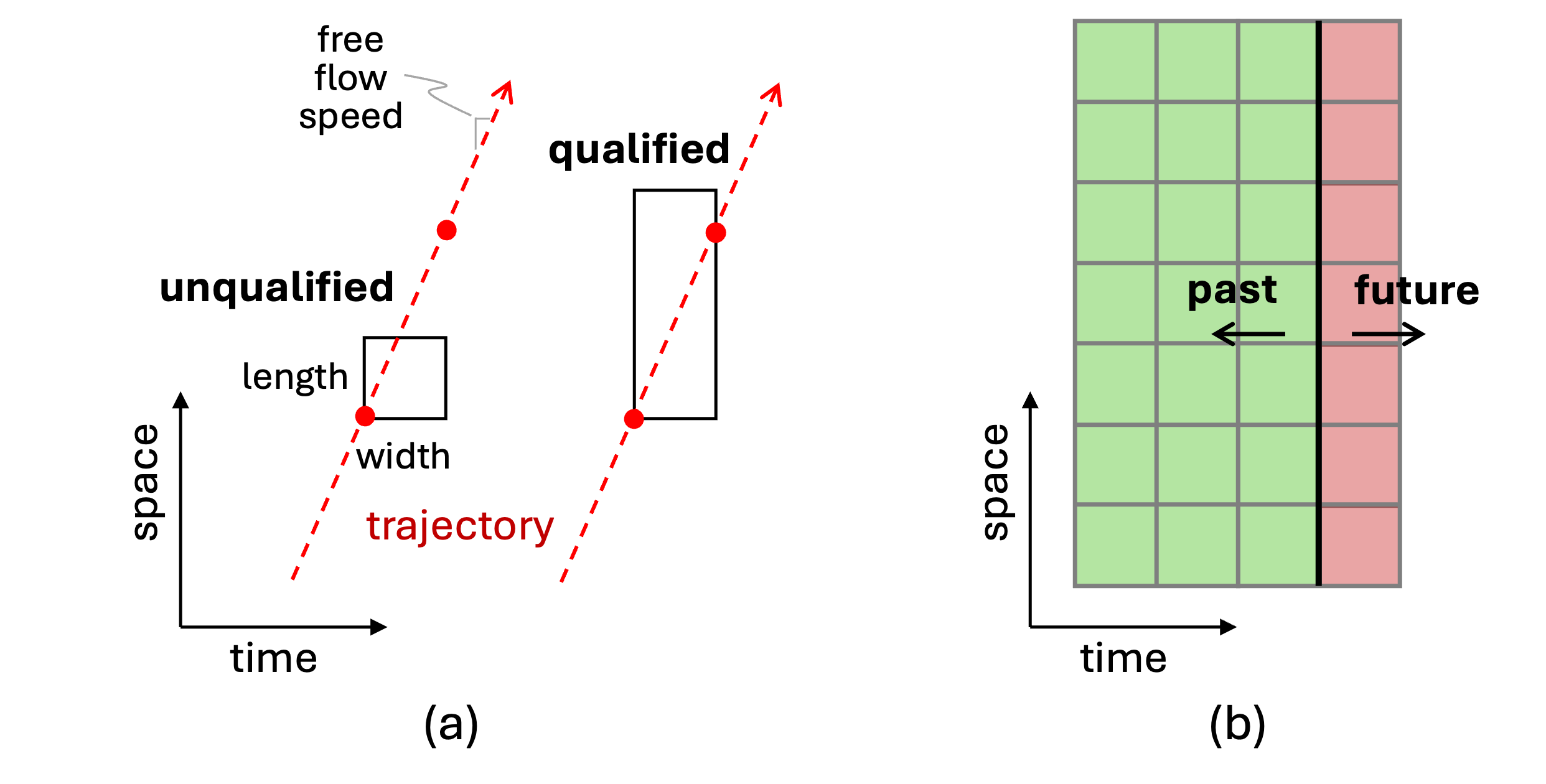}
    \caption{(a) Current constraints for cell size. The unqualified cell is unable to catch every flow (e.g., data point in a trajectory) in the given time interval; (b) Cell-based traffic state prediction. There should be three versions of the time-space diagrams, i.e., speed, flow, and density.}
    \label{fig:prediction}
\end{figure}

As pointed out by most of the current research, traffic state estimation and prediction are the foundation of the successful execution of wave suppression.
Despite the importance, most of the current works do not incorporate state-of-the-art approaches.
For instance, the current MPC-based VSL strategies heavily rely on the prediction of cell-based models such as METANET in \cite{VSL2005,VSL2007} and CTM in \cite{VSL2017,VSL2017b}.
In addition to the inherent limitations of model-based approaches (such as oversimplification), the resolution requirements of those cell-based models seem to be overlooked.
Specifically, the hard constraint for those models that guarantees conservation laws (i.e., all flow falls in a cell at least once) is
$L_\text{C} > v_\text{f}T_\text{C}$, 
where $v_\text{f}$ is the free-flow speed, and $L_\text{C}$ and $T_\text{C}$ are the length (in space) and width (in time) of a cell, respectively (Figure \ref{fig:prediction}(a)).
The typical free-flow speed on a freeway is 120 km/h. Therefore, for a cell with a 10-second time width (typically used by METANET), the corresponding length must exceed 300 meters, which may be too coarse to effectively capture traffic dynamics with the wave speed of around 16 km/h in practice, particularly in multiple stop-and-go waves \cite{He2024}.

To overcome the constraints on granularity, traffic state estimation approaches, such as the adaptive smoothing method (ASM), have been demonstrated to be both meaningful and effective \cite{Hegyi2013}.
It has yet to be seen that more advanced traffic state estimation approaches \cite{Seo2017}, such as deep learning-based approaches, are employed.
In solving the time-space diagram refinement problem, \cite{He2024} has demonstrated that the constraint on the aspect ratio of the cells can be effectively relaxed by introducing a multiple linear regression model. 
Moreover, the time-space discrete cell can be treated as a pixel containing information on speed, flow, and density, which is fundamentally similar to an image (Figure \ref{fig:prediction}(b)).
Convolutional Neural Networks (CNNs) are particularly invented to handle such objects, and they have demonstrated effectiveness in traffic prediction \cite{Ma2017}.
Physics-informed machine learning, which integrates physical principles and domain knowledge into machine learning models to improve their accuracy, generalization, and interpretability \cite{Karniadakis2021}, is also a promising future direction.
These advanced machine learning-based predictions can be effectively supported by the rich and diverse high-fidelity traffic data recently made available, such as highD \cite{Krajewski2018}, ZenTraffic \cite{Dahiyal2020}, and I-24 MOTION \cite{Gloudemans2023}.

It is worth noting that, beyond accurate prediction of stop-and-go wave propagation, inflow or demand prediction is also critical for effective suppression, as reported by \cite{Youngjun2015}. However, this aspect appears to be largely overlooked in most existing studies, which often assume solely a constant and typically high inflow that sustains stop-and-go waves.

\subsection{Robustness to Randomness}

Most of the theory-based strategies or analytical models rely on a theory that can describe traffic dynamics, such as KWT at the macroscopic level and CF models at the microscopic level.
Although these fundamental theories play an important role in modeling and analysis, the stochastic nature of traffic indicates that these oversimplified deterministic models, which basically describe an equilibrium or average behavior, may not be sufficient to support practical applications. Therefore, it is necessary to adapt existing strategies to encompass stochastic extension of the current fundamental theories.
For instance, \cite{Laval2013} extended KWT by incorporating stochastic FDs and stochastic initial conditions, which can be employed to enhance the current KWT-based strategy and analysis, such as considering stochastic factors when tuning SPECIALIST \cite{VSL2008} and enhancing the theory of JAD \cite{JAD2013,JAD2024b}. 
Similarly, stochastic CF models, such as those in \cite{Xu2020,Tian2021}, can be used to derive the conditions for stop-and-go wave suppression and the prevention of secondary waves \cite{JAD2015,JAD2020b}.
Among the existing VSL and JAD works, \cite{JAD2025} provides a valuable example, as it attempts to randomize a deterministic CF model and incorporate it into the existing JAD framework for wave propagation estimation.
A similar technique could also be applied in other aspects of the overall framework of stop-and-go wave suppression.

It is also true for MPC-based strategies.
The requirement of MPC for a well-defined predictive model makes it challenging to incorporate the stochastic nature of traffic dynamics.
To address this issue, various control schemes might be helpful. For instance, robust control can be employed to handle the uncertainty in the length and speed of a moving queue during propagation. Stochastic MPC can also be considered to enhance existing MPC-based strategies.
To mitigate the difficulty of traffic dynamics modeling, data-driven predictive control (DPC) could be a viable option, reducing the reliance on (macroscopic) traffic flow models in MPC-based strategies. DPC is a control scheme that learns controllers directly from data, eliminating the need for extensive system knowledge or explicit model identification, which presents a key advantage over traditional MPC \cite{Schwenzer2021, Verheijen2023}.

As a rapidly developing field, RL warrants further exploration for wave suppression tasks, given its inherent capability to handle stochasticity and non-stationary dynamics.
To date, only one RL-based strategy has been identified, and it is only for VSL \cite{VSL2022}. There is significant potential for follow-up research, such as adopting a microscopic perspective or introducing RL into JAD, as \cite{JAD2025} introduced MPC into JAD. In this context, it would be valuable to consider stochastic factors, such as heterogeneous driving behavior and human drivers' imperfection \cite{Laval2010}.

In testing a strategy, although simulation is unavoidable most of the time, the more realistic the simulation is, the more convincing the strategy will be and the closer it will be to practical deployment.
In particular, accurately reproducing the capacity drop—preferably aligned with the findings of \cite{Yuan2016} on the relationship between queue discharge rate and in-queue speed—as well as the instability and breakdown phenomena at bottlenecks, is essential.
Those stochastic (particularly microscopic) models, such as those advanced models mentioned in \cite{Chen2023,Zhang2024Car}, are expected to be employed, while most of the current strategies were tested in a deterministic or aggregated model such as IDM, METANET, and CTM.
It is worth mentioning that many data-driven traffic flow models have recently emerged \cite{Chen2023,Zhang2024Car}. Since these models are independent of any specific mathematical framework and incorporate randomness, using them to test a strategy can lead to a promising validation \cite{JAD2017,JAD2025}.


\subsection{Scenarios for Strategy Validation}

\begin{figure*}[htbp]
    \centering
    \includegraphics[width=0.85\linewidth]{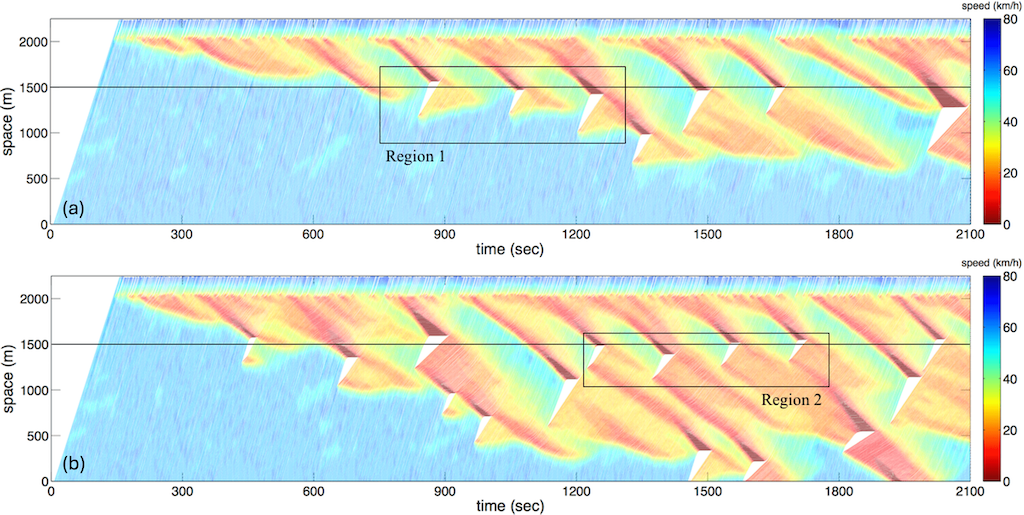}
    \caption{JAD performance in a more challenging multiple-wave scenario (sourced from \cite{JAD2017}). Both (a) and (b) share the same inflow. Fewer waves are generated in (a) by setting a lower slowdown probably at the bottleneck, while more in (b).}
    \label{fig:multiple_wave}
\end{figure*}

As shown in Figure \ref{fig:wave}, it is quite common to observe that multiple and periodic stop-and-go waves appear on the freeway at the same time, and therefore, it is necessary to evaluate the strategies in a more practical and common scenario. 
Unfortunately, as shown in Table \ref{tab:Summary}, many strategies are validated in single-wave scenarios, which are relatively simple compared to the real-world observations.
Single-wave scenarios are well-suited for initially testing the strategies, while multiple-wave scenarios allow for a more thorough assessment of the strategies in a setting closer to real-world conditions. 
Obviously, multiple-wave scenarios present greater challenges (Figure \ref{fig:multiple_wave}) \cite{JAD2017,JAD2024b}.
For the cell-based MPC strategies, the model resolution, i.e., the cell size, may not be sufficiently fine to distinguish multiple waves. However, it remains uncertain whether the current cell size is sufficient to capture multiple waves, or what level of granularity would be appropriate \cite{Wang2022}. 
Even for the SPECIALIST, although it was designed for the single (or isolated) wave scenario, it still deserves a test in a more realistic scenario such as the one shown in Figure \ref{fig:wave}(d) \cite{VSL2015}. Then, new problems can be uncovered, inspiring new research questions and corresponding solutions.



Given the current abundance of real-world traffic data \cite{Li2020}, it is now an opportune time to establish trustworthy simulation scenarios for various transportation studies, including both VSL and JAD strategies.
For instance, boundary conditions, i.e., vehicle types, speeds, and timestamps at the entry and exit of the simulated freeway stretch, can be extracted and fed into a traffic simulation model or platform.
The trustworthiness of the simulation can be further improved if the simulation models, such as CF models, LC models, and even the origin-destination matrix, could be calibrated using vehicle trajectory data collected from the same freeway segment.



\subsection{Field Test and Practical Deployment}\label{sec:Field}

\subsubsection{Variable Speed Limit}

\begin{figure}[htbp]
    \centering
    \includegraphics[width=\linewidth]{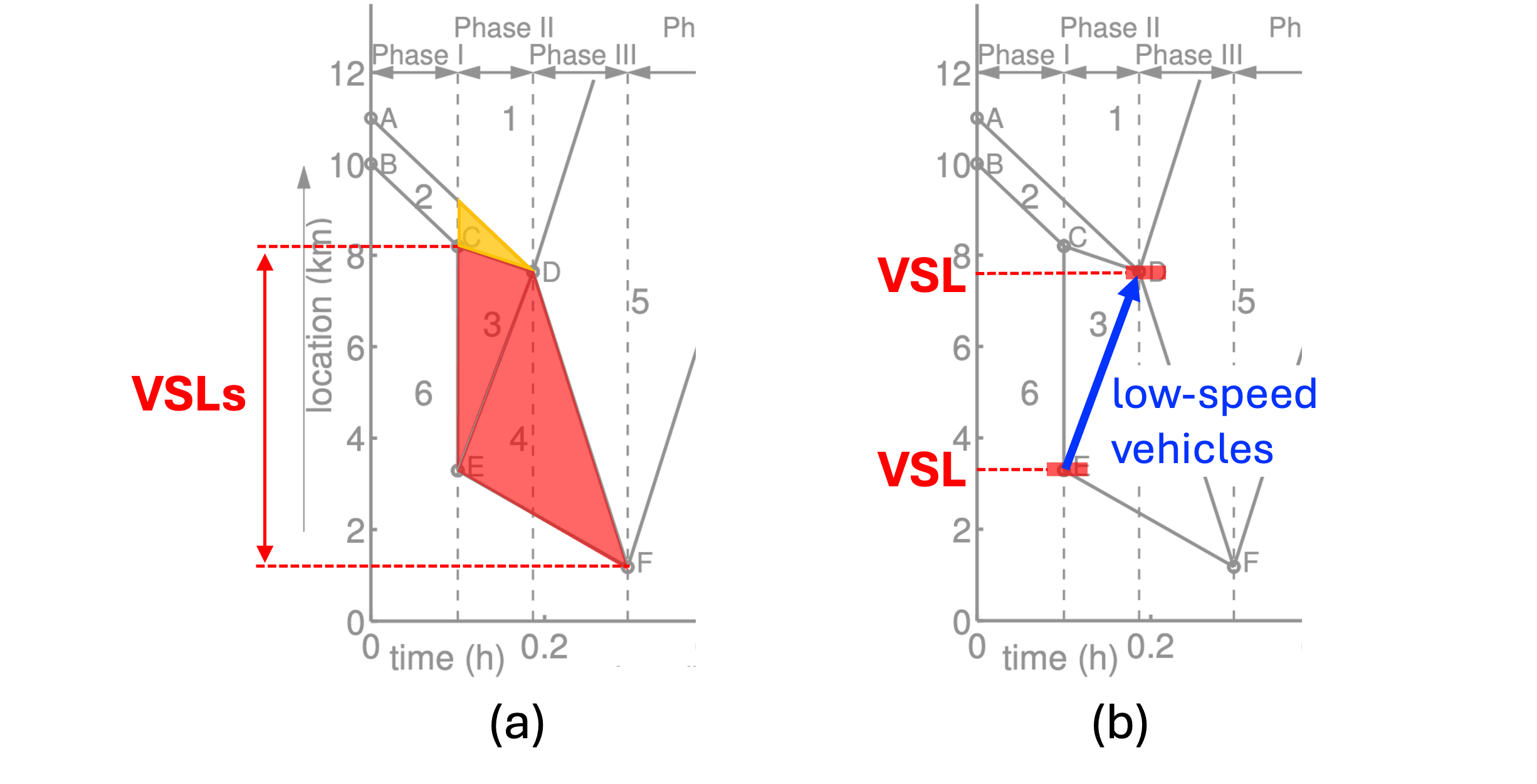}
    \caption{Two ways to disseminate speed limits (The background image is sourced from \cite{VSL2010}): (a) One-shot plan generated by SPECIALIST, where the VSL displayed across multiple VMS is scheduled in advance based on the time-space plan shown in the figure. Note that controlling the yellow region is not theoretically required but is practically preferred, particularly to extend the speed limits into area 2 downstream of line CD. It would feel illogical to drivers if the speed limit is lifted just as they enter the jam. (b) bi-VMS configuration.}
    \label{fig:control}
\end{figure}

To date, \cite{VSL2010} might still be the only real-world test for stop-and-go wave suppression strategies.
Besides policy-related challenges, the significant infrastructure investment is also a major obstacle. 
In \cite{VSL2010}, over 20 gantries with VMS were installed along a 14 km freeway stretch.
Once a qualified target wave is detected and identified, the system determines the complete control plan in one shot and then disseminates the VSL according to the predetermined plan (Figure \ref{fig:control}(a)). 
As with most open-loop control strategies, the inability to make real-time adjustments during execution limits operational flexibility.


At present, the VSL strategies require the installation of a series of gantries and VMS for support. Referring to JAD strategies, it may be feasible to propose a low-cost bi-VSL strategy, in which the upstream VSL reduces the speed limit, while the downstream one is responsible for setting the speed back to normal (Figure \ref{fig:control}(b)). The bi-location VSL might not only be more practical but also sufficient to set a low-speed region to prevent stop-and-go wave propagation. 
Still, it is worth noting that—similar to JAD, as discussed at the end of Section \ref{sec:JAD}—the stability behind the blue line in Figure \ref{fig:control}(b) remains unclear and warrants further investigation.


\subsubsection{Jam Absorption Driving}

Compared to VSL, JAD has not yet undergone any field testing, despite offering greater advantages in terms of both investment costs and implementation flexibility.
However, we have the following three empirical observations or advancements, which could be \textit{strong evidence} that shows JAD is practically feasible and also effective.

\begin{figure}[htbp]
    \centering
    \includegraphics[width=\linewidth]{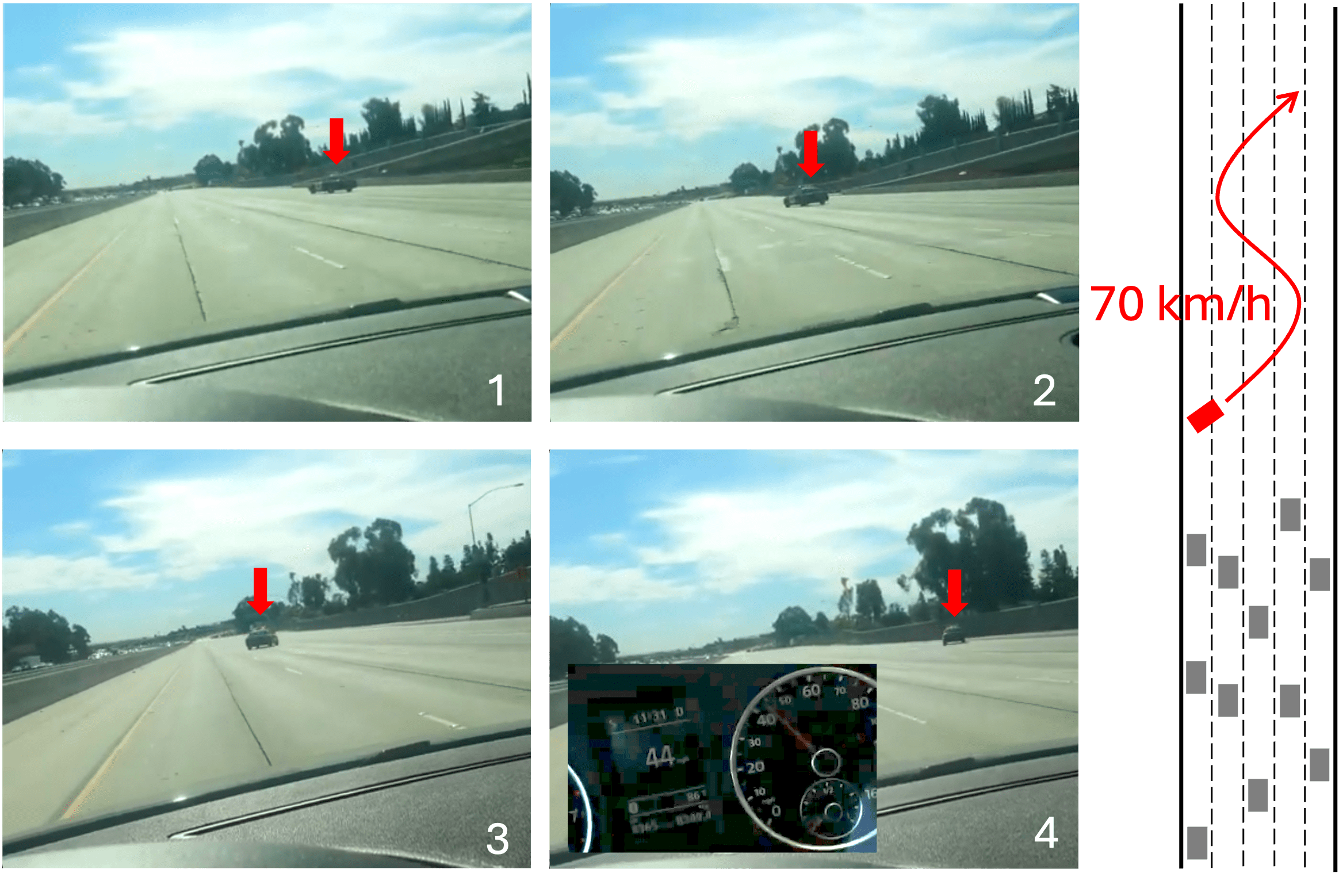}
    \caption{Police car swerving at high speed on a freeway, California, USA (Snapshots from the video in \cite{Carpenter2025video}).}
    \label{fig:swerving}
\end{figure}

First, the behavior similar to JAD has occurred in the real world. 
As shown in Figure \ref{fig:swerving}, a police car was recorded swerving on the freeway in California, USA, to control the following traffic at a high speed.
In Figure \ref{fig:swerving}, it was 70 km/h, which is higher than the lowest VSL at 60 km/h required by many VSL studies. The real-world police car swerving highlights the feasibility of JAD.
Although the goal of the police car in this example might not be jam absorption, it greatly inspires a potential real-world implementation of JAD.
In addition, we are still not sure how often such swerving behavior happens in practice, but it is not difficult to find similar videos on the Internet.

Regarding the traffic stability behind the JAD vehicle, unfortunately, no data is currently available for the traffic conditions behind the police car.  It would be interesting to see whether the upstream flow remained stable, as this could offer important insights into the feasibility and effectiveness of single- or few-vehicle-based JAD.

Considering such an execution method of police car intervention, a more practice-oriented research question would be to start with direct instructions for a police car on where and when to begin and at what speed to drive, and then work backward to determine the necessary information and estimates, rather than assuming possession of all information and the ability for a JAD vehicle to appear anywhere and anytime. 
This strategy would be rather ideal if we could suppress a stop-and-go wave before it propagates to a location where a police car can be deployed, i.e., we may not have to suppress it immediately after a stop-and-go wave appears. 
This might be the most direct way to deploy JAD strategies so far, requiring neither additional investment nor the involvement of CAVs.

\begin{figure}[htbp]
    \centering
    \includegraphics[width=\linewidth]{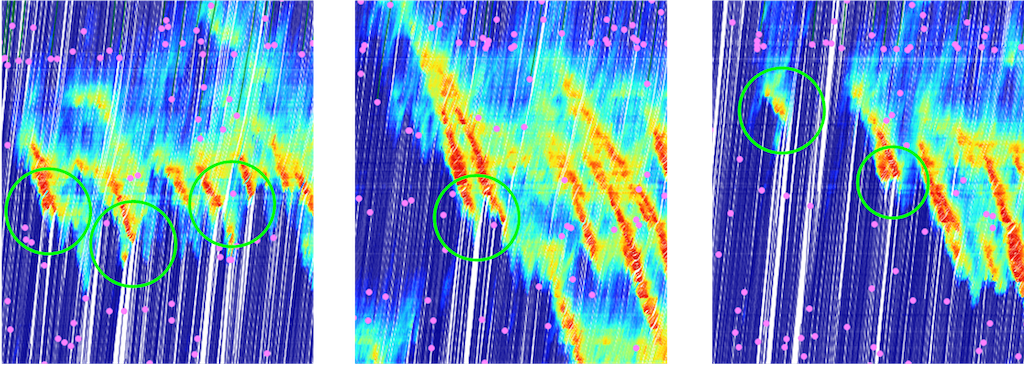}
    \caption{Empirical observations suggest that maintaining an appropriate gap from the leading vehicle when joining a stop-and-go wave may effectively suppress its propagation without triggering secondary waves. The regional screenshots were extracted from time-space diagrams of vehicle trajectories, which were plotted using data collected from the Hanshin Expressway Route 11 (Ikeda Route) in Osaka City, Japan \cite{Dahiyal2020}.}
    \label{fig:JAD_data}
\end{figure}

Second, empirical data suggest that maintaining an appropriate gap from the leading vehicle when joining a stop-and-go wave could be effective in suppressing it, as shown in Figure \ref{fig:JAD_data}.
Although one may argue that this observation is related to the nature drop in demand, it still implies the potential feasibility of JAD, in particular from the microscopic perspective. The key lies in identifying the applicable conditions under which JAD can be effectively implemented.
Therefore, it suggests another promising research direction, which is to extract similar scenarios from the existing extensive empirical trajectory datasets and analyze them to identify the conditions under which such behavior could contribute to wave suppression, ultimately proposing an empirical observation-based JAD strategy or even other CAV-based strategies beyond JAD.

Third, CAV should not be neglected. The recent field experiment of traffic control with 100 CAVs is insightful \cite{Lee2025}, despite the high cost in both investment and workload.
It particularly highlights the possibility and feasibility of intervening traffic flow through dedicated vehicles even in large numbers, implying the potential of not only JAD strategies but also those VSL-JAD integrated strategies.


\section{Concluding Remarks}\label{sec:Conclusion}

\begin{figure}[htbp]
    \centering
    \includegraphics[width=0.94\linewidth]{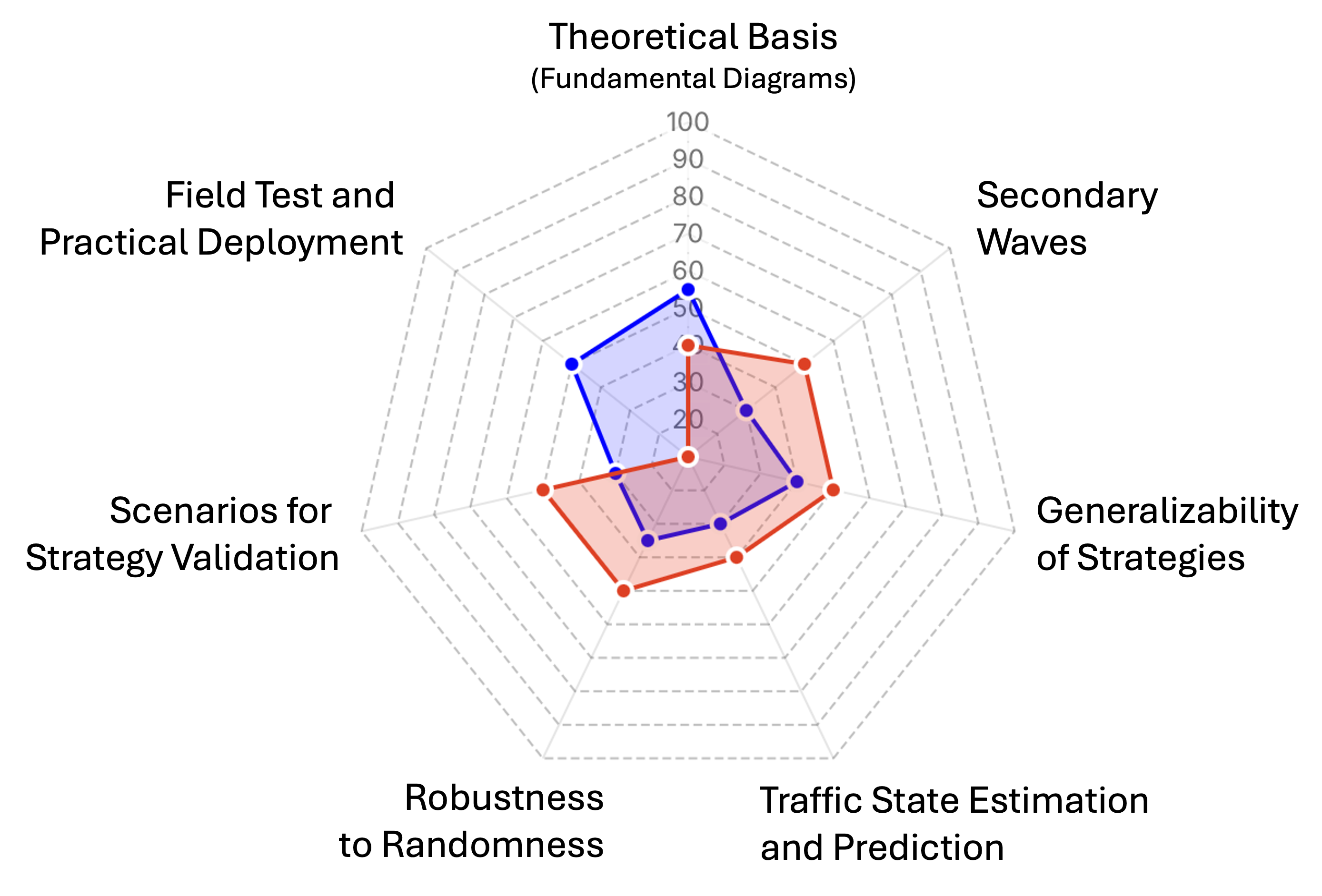}
    \caption{Authors' subjective assessment of VSL vs. JAD development.}
    \label{fig:radar}
\end{figure}

Stop-and-go waves are the common pattern of congestion on freeways/highways worldwide.
With their severe adverse effects on traffic efficiency, driving safety, and emissions, stop-and-go waves persistently affect human society on a broad time-space scale.
Over decades of research and development, a variety of measures and strategies have emerged to suppress the propagation of stop-and-go waves, including mainly VSL and JAD strategies.
Unfortunately, despite important similarities, such as the shared goals and underlying logic, a disconnection between the two areas is observed.
On the occasion of the 20th anniversary of the first stop-and-go wave suppression strategy \cite{VSL2005} being proposed, this review summarizes the progress of both VSL and JAD areas in detail, pinpoints the strengths and weaknesses of each (Figure \ref{fig:radar}), and seeks to bridge the two areas, identify research opportunities, and streamline research directions.

Learning from JAD strategies, we recommend that the VSL area to place more focus on thorough validation using more realistic (e.g., stochastic) microscopic traffic simulations and in more realistic scenarios. The secondary wave issue (particularly type-II and type-III), which has been widely reported in the JAD area and is highly likely to exist in the VSL area, urgently needs to be tested and addressed.
To enhance traffic estimation and prediction, more advanced techniques, such as deep learning, could be incorporated. To increase the generalization of the strategies, it is beneficial to consider more diverse traffic patterns, as well as introduce the latest AI tools (e.g., LLMs) to reduce the workload of fine-tuning. 
To reduce the need for constructing continuous, high-cost gantries for VSL strategies, a lightweight bi-VSL strategy could be proposed, drawing inspiration from JAD strategies.

Compared to VSL strategies, the theoretical basis underlying the design of JAD strategies is less well developed.
This may be the main reason why the secondary wave issue has been inadequately addressed, and some research may even be biased. 
The real-world test in \cite{VSL2010} provides strong evidence for the effectiveness of VSL strategies. Unfortunately, no similar field test for JAD has been reported.
Since they share similar logic, JAD should also work similarly to VSL. Although it is not directly related to stop-and-go wave suppression, the observation of a police car swerving on a freeway at high speeds provides us with great confidence that JAD is also practically feasible.
Making the current requirements (i.e., model input) for JAD implementation more practical is a valuable research effort that we can make before a real-world test opportunity arises.

Despite the disconnection, we are also pleased to observe that the two areas are merging, particularly at the technical level, i.e., the role of vehicles in intervening traffic is being strengthened with the rapid development of CAV and V2I technologies, further highlighting the importance of the synergy between the two areas.

Unlike other areas of transportation research that center on (e.g.) understanding mechanisms or identifying patterns, freeway traffic control directly interacts with the traffic system, placing exceptionally high demands on the robustness and adaptability of the models and strategies. 
Given the similarities in research questions and goals, it might be worth exploring a \textit{unified framework} that decouples modeling from implementation.
Specifically, at the theoretical level, we may focus on stop-and-go wave suppression-oriented traffic flow optimization, through, e.g., traffic-based optimal control with the goal of minimizing travel time or suppressing the waves.
In implementation, choosing between VSL and JAD can be guided by their respective strengths and limitations, depending on the practical conditions, with the aim of achieving the most effective outcome.
Such a plug-in modeling-implementation structure enhances the overall framework's generality and flexibility.





\bibliographystyle{IEEEtran}
\bibliography{library}

\vspace{-1cm}

\begin{IEEEbiography}[{\includegraphics[width=1in,height=1.25in,clip,keepaspectratio]{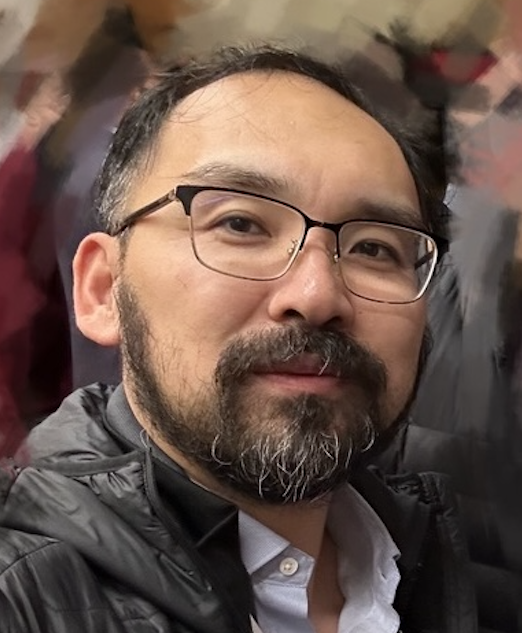}}] 
{Zhengbing He} (M'17-SM'20) received the Bachelor of Arts degree in English language and literature from Dalian University of Foreign Languages, China, in 2006, and the Ph.D. degree in systems engineering from Tianjin University, China, in 2011. He was a Post-Doctoral Researcher and an Assistant Professor with Beijing Jiaotong University, China. From 2018 to 2022, he was a Full Professor with Beijing University of Technology, China. Presently, he is a Research Scientist with the Laboratory for Information and Decision Systems (LIDS), Massachusetts Institute of Technology, USA. 

His research stands at the intersection of transportation, systems engineering, and artificial intelligence, with a focus on topics such as data-driven modeling and intelligent vehicle-enabled congestion solutions, learning-based sensing and prediction of traffic congestion and travel demand, and sustainability-oriented optimization in transportation. He has published more than 160 academic papers, with total citations exceeding 6,000. He was listed as World’s Top 2\% Scientists. He is the Editor-in-Chief of the Journal of Transportation Engineering and Information (Chinese). Meanwhile, he serves as a Senior Editor for IEEE TRANSACTIONS ON INTELLIGENT TRANSPORTATION SYSTEMS, Deputy Editor-in-Chief of IET Intelligent Transport Systems, a Handling Editor for Transportation Research Record, and an Editorial Advisory Board Member for Transportation Research Part C. His webpage is https://www.GoTrafficGo.com.
\end{IEEEbiography}

\vspace{-1cm}

\begin{IEEEbiography}[{\includegraphics[width=1in,height=1.25in,clip,keepaspectratio]{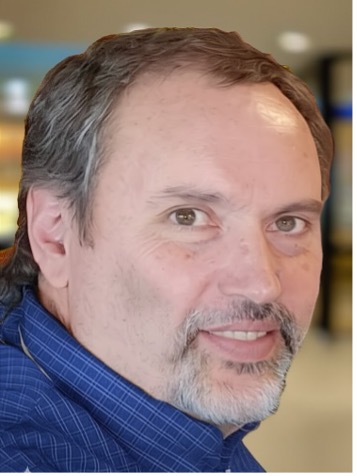}}] 
{Jorge A. Laval} is a Professor at Georgia Tech in the School of Civil and Environmental Engineering. He obtained his PhD in Civil Engineering from the University of California, Berkeley under the supervision of Carlos Daganzo. His main research focus is understanding urban traffic congestion to devise improved control methods, combining aspects from traffic flow theory, complex systems theory and machine learning. Recently, he has made important contributions towards understanding the significant challenges that machine learning methods have when it comes to urban network control, and towards improved autonomous vehicle control models that promote traffic stability. Other contributions include methods for estimating the macroscopic fundamental diagram of urban networks, and stochastic car-following models that allow the statistical inference of parameters.
\end{IEEEbiography}

\begin{IEEEbiography}[{\includegraphics[width=1in,height=1.25in,clip,keepaspectratio]{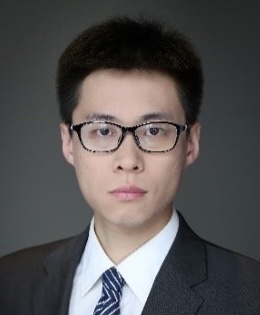}}] 
{Yu Han} is an Associate Professor at the School of Transportation, Southeast University. He obtained his BSc (2011) from Harbin Institute of Technology, MSc (2013) from Beijing Jiaotong University, and PhD (2017) from Delft University of Technology. His primary research interests include developing traffic control strategies at both macro and microscopic levels through interdisciplinary approaches, such as traffic flow theory, control theory, optimization, and artificial intelligence. He has authored more than 30 publications in peer-reviewed journals. He contributed to several major projects focused on algorithm development for intelligent highway systems in multiple provinces across China. He was awarded the China Highway and Transportation Society Science and Technology Award (Top Prize) in 2023 and the Jiangsu Province Science and Technology Award (First Prize) in 2024.
\end{IEEEbiography}

\vspace{-2cm}

\begin{IEEEbiography}[{\includegraphics[width=1in,height=1.25in,clip,keepaspectratio]{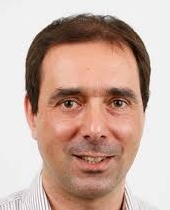}}]{Andreas Hegyi} received the M.Sc. degree in electrical engineering and the Ph.D. degree from TU Delft, The Netherlands, in 1998 and 2004, respectively. He is currently an Assistant Professor with TU Delft. He is the author or coauthor of over 100 papers. His research interests include traffic flow modeling and control, connected and cooperative vehicles, traffic state estimation, and traffic data analysis. He is a member of IEEE-ITSS and IFAC-CTS. He has served as the Program Chair for the IEEE-ITSC 2013 Conference, the General Chair for the IXth TRISTAN Symposium 2016, and an IPC member for various other conferences. He is an Associate Editor of IEEE Transactions on Intelligent Transportation Systems.
\end{IEEEbiography}

\vspace{-2cm}

\begin{IEEEbiography}[{\includegraphics[width=1in,height=1.25in,clip,keepaspectratio]{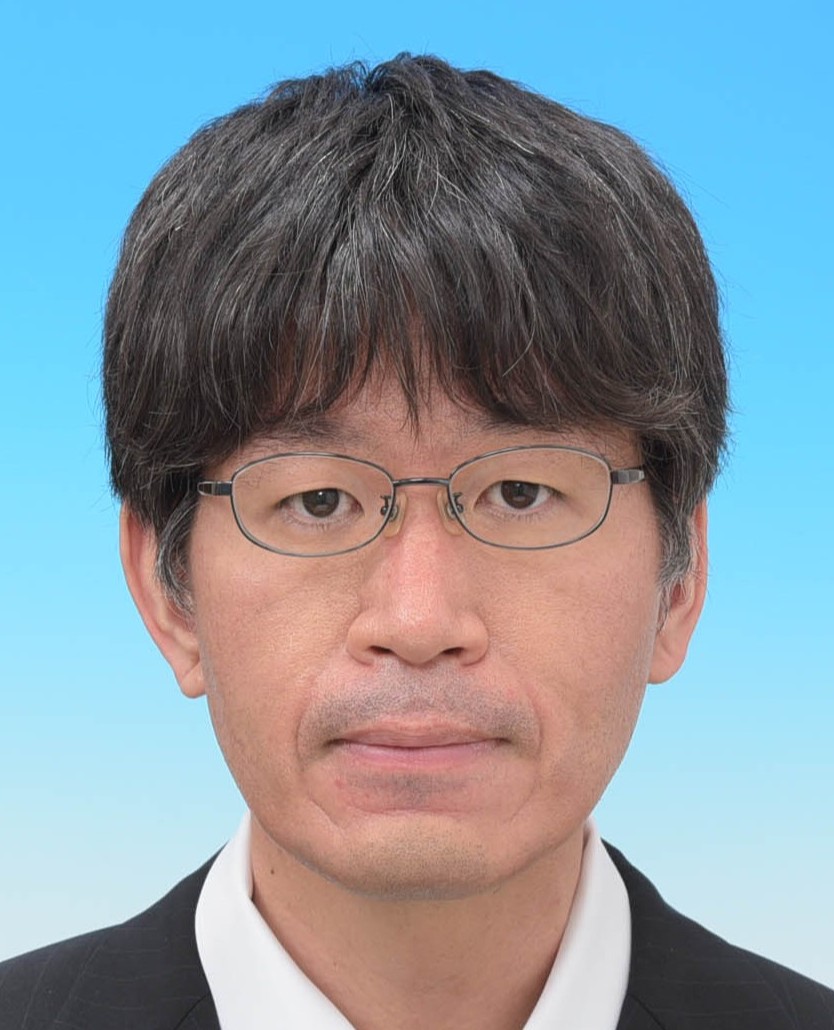}}] 
{Ryosuke Nishi} is an Associate Professor at Department of Mechanical and Physical Engineering, Faculty of Engineering, Tottori University, Japan. He received his B.S., M.S. and Ph.D. in Engineering from the University of Tokyo in 2007, 2009 and 2012, respectively. He worked as a project researcher at the University of Tokyo from April 2012 to October 2012 and at National Institute of Informatics, Japan from November 2012 to November 2014. He has been working at Tottori University since December 2014. He is interested in the collective dynamics of vehicular traffic flow, especially mitigation of freeway congestion with an extremely small number of vehicles.
\end{IEEEbiography}

\vspace{-2cm}

\begin{IEEEbiography}[{\includegraphics[width=1in,height=1.25in,clip,keepaspectratio]{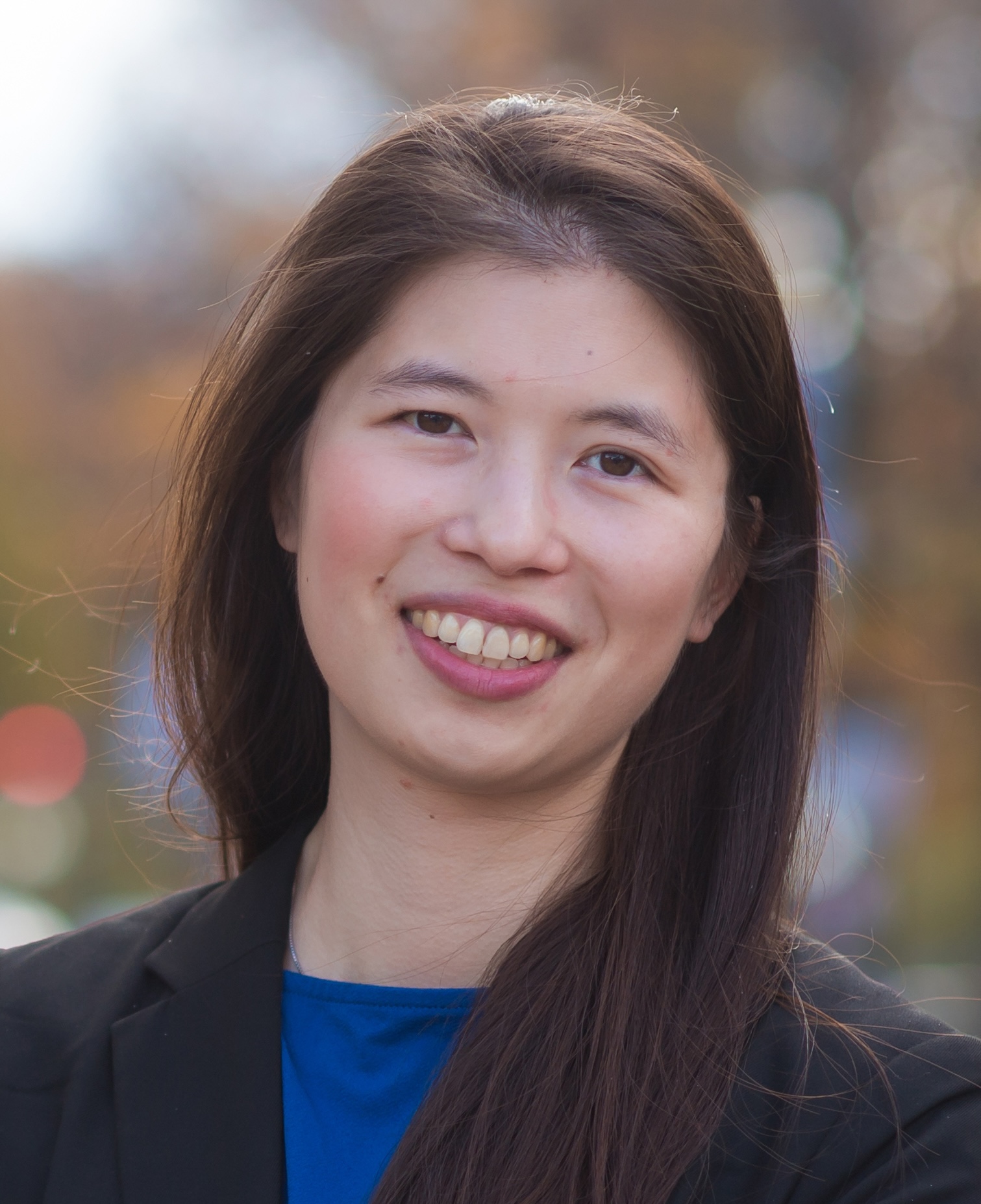}}]{Cathy Wu} is the Thomas D. and Virginia W Cabot Career Development Associate Professor of Civil and Environmental Engineering at MIT and holds additional appointments in LIDS and IDSS. She holds a Ph.D. from UC Berkeley and B.S. and M.Eng. from MIT, all in EECS, and completed a Postdoc at Microsoft Research. Her interests are at the intersection of machine learning, control, and mobility. Her recent research focuses on how learning-enabled methods can better cope with the complexity, diversity, and scale of control and operations in mobility systems. She is interested in developing principled computational tools to enable reliable decision-making in sociotechnical systems.
\end{IEEEbiography}

\end{document}